\documentclass[aps,prb,twocolumn]{revtex4}

\usepackage{soul}
\usepackage{color}
\usepackage{graphicx}
\usepackage{enumitem}
\usepackage{amsfonts}
\usepackage{amsmath,latexsym,amsthm}
\usepackage{bbold}
\begin{document}

\title{{Black holes, fast scrambling and the breakdown of the
equivalence principle}}

\author{Zhi-Wei Wang${}^{1,\ast}$, Saurya Das${}^{1,\dagger}$,
and Samuel L.\ Braunstein${}^{2,\ddagger}$}

\affiliation{${}^1$Theoretical Physics Group, 
Department of Physics and Astronomy,
University of Lethbridge, 4401 University Drive, Lethbridge,
Alberta T1K 3M4, Canada}
\affiliation{${}^2$Computer Science, University of York, York YO10 5GH,
United Kingdom}
\affiliation{${}^\ast$zhiweiwang.phy@gmail.com}
\affiliation{${}^\dagger$saurya.das@uleth.ca}
\affiliation{${}^\ddagger$sam.braunstein@york.ac.uk}

\begin{abstract}
\noindent 

Under reasonable assumptions, black holes have been argued to form
firewalls, burning up anything crossing their horizons. This argument
finds that a firewall would appear very late in a black hole's lifetime,
when Hawking radiation has caused the horizon to shrink to one-half its
original area. For stellar-mass black holes, this process surpasses the
universe's current age and so no such black hole would currently possess
a firewall. However, black holes have recently been conjectured to
scramble their interior degrees-of-freedom, with a scrambling time scale
comparable to the time it takes light to travel a Schwartzschild radius'
distance. We prove that local observers will already experience a
firewall from the scrambling time onwards after the black hole's
formation. Here `local' means that the observer couples to fewer than
one-half the black hole's total interior `qubits.' Indeed, for observers
to fail to be local in this manner, it would mean that they couple to
more `qubits' within such black holes than exist in all the stars of the
observable universe. Therefore we find that if black holes are indeed
fast scramblers, then every astrophysical black hole in the universe
will already have a fully developed firewall for any local physical
process.

\end{abstract}

\maketitle

\section*{I. Introduction}

{
 The equivalence principle of classical general relativity
tells us that if we are in free fall, we do not feel the effects of
gravity locally.\cite{Einstein1907} Traditionally, this principle leads
to the expectation of `no-drama' when crossing the horizon into a large
black hole. In other words, an observer, small compared to the size of
the black hole, crossing the event horizon should not detect anything
unusual. Only when they approach too close to the central singularity,
where tidal forces become extreme will the effects of gravity become
apparent. From the viewpoint of the equivalence principle, therefore,
crossing the horizon should not manifest any violent phenomena. 

This behavior is challenged by quantum effects originating from
degrees-of-freedom near the horizon that lead to the emission of thermal
radiation, first famously predicted by Hawking.\cite{Hawking76} Quantum
mechanical black holes are predicted to (slowly) evaporate away, yet
according to the classical equivalence principle, the information
(carried say by a freely-falling observer) that has fallen into a black
hole can make its way deep into the interior where it cannot participate
in the evaporation process. Thus, the Hawking radiation carries away
mass from the black hole, but not the information it contains. Were this
situation to continue until the black hole had completely evaporated we
would have a paradox since quantum mechanics relies on the preservation
of information.\cite{Hawking76} This is the essence of the famous black
hole information paradox and is a direct consequence of the clash
between the expectations of no drama from the classical equivalence
principle and the quantum effects leading to Hawking radiation and black
hole evaporation. (See, e.g., Refs.~\onlinecite{Davies1978,Preskill1992,Wald2001,Harlow2016,Marolf2017,Polchinski2017Review}, for a more comprehensive
discussion of the paradox and potential loopholes.)
}

Outside a large black hole, physics is largely well understood. For
stationary observers, an outgoing flux of radiation is observed. At
spatial infinity, this Hawking radiation has a temperature that scales
as $T_H\propto O({1}/{M})$ for a black hole of mass
$M$.\cite{Hawking76} According to the `membrane paradigm,' a simple
thermodynamic argument suggests that stationary observers closer to the
black hole should see a blue-shifted flux of this radiation, reaching a
universal temperature of roughly one Planck energy (taking the Boltzmann
constant as unity) when the stationary observer is roughly one Planck
length from the horizon.\cite{Thorne10} Similarly, a freely-falling
observer, sees an outward flux of radiation for distances larger than
$O(3M)$.\cite{Davies76} However due to quantum field renormalization
effects, this flux reverses itself for observers nearer the horizon, and
exactly vanishes for infalling observers as they pass the
horizon.\cite{Davies76} 
%
%
This latter result at the horizon is often interpreted as following from
the equivalence principle.\cite{Susskind93}

However, what about the physics inside the horizon of a black hole?
Naively, the equivalence principle should continue to hold and a small
infalling observer  (and hence due to locality, one coupling to a
limited number of degrees-of-freedom in its neighborhood within the
black hole) should continue to notice nothing special until they are
torn apart by gravitational stresses as they approach the singularity.
On the contrary, calculations based on quantum models suggest that an
infalling observer will observe high-energy quanta near the horizon of a
black hole which is older than the Page time (an `old black hole') -- when
the horizon area has shrunk by a factor of
two.\cite{braunstein2009,almheiri2013} This led to the proposal that
the horizon of an old black hole should be replaced by a
firewall.\cite{braunstein2009,almheiri2013}

The original proof of the firewall paradox\cite{almheiri2013} required
the infalling observer to extract enough information from the already
present outgoing Hawking radiation before reaching the horizon. However,
calculations based on quantum computation show that the time to extract
this information is generally longer than the lifetime of the black
hole, which forms a potential loophole to the original firewall claim.
\cite{harlow2013} 


{ In addition to these approaches, a number of thorough
reviews and discussions of the firewall paradox and black hole
information puzzle can be found in
Refs.~\onlinecite{Harlow2016,Polchinski2017Review}. These works explore
the conceptual underpinnings of black hole complementarity, entanglement
at the horizon, and quantum computational aspects of information
retrieval from black holes. Furthermore, recent progress in
understanding black hole interiors via the quantum extremal surfaces or
`islands' program has also provided new insights into how entanglement
wedges and late-time radiation might resolve the firewall
paradox.\cite{Almheiri2019,Penington2019,Almheiri20} Such approaches
emphasize the deep connections between the geometry of spacetime and the
quantum information-theoretic entanglement structure of black hole
states.

For a general physics audience, it can be useful to frame the firewall paradox 
in the context of the long-standing black hole information problem: 
the apparent conflict between unitarity (i.e., quantum information 
cannot be destroyed) and the classical expectation that information 
falling into a black hole is lost behind the event horizon. 
The firewall scenario posits an extreme resolution of this puzzle, 
suggesting that near-horizon quantum correlations break down for old black holes, 
leading to high-energy quanta at or just behind the horizon, thus violating the usual `no-drama' experience.

The mechanism behind the firewall can be understood in
simple terms. Assuming the original black hole is created in a pure
quantum state, unitarity tells us that the total Hawking radiation from
a completely evaporated black hole will be likewise pure. This overall
purity requires perfect entanglement between the late (post-Page time)
and early (pre-Page time) Hawking
radiation.\cite{braunstein2009,page1993information} Consequently, the
Page-time-aged black hole that evaporates into this late radiation, must
have been in a maximally mixed state; implying it has a firewall.

Yoshida, for example, considers a scenario in which $k$ qubits of
matter, maximally entangled with an external reference, are thrown into
a black hole at its Page time.\cite{Yoshida2019} He shows that the early
radiation is actually unentangled with the subsequent $k$ qubits of
Hawking radiation, hence breaking the usual early-late entanglement
requirement for the firewall argument.\cite{Yoshida2019} He claims that
this provides a resolution for the paradox without a firewall
appearing.\cite{Yoshida2019}

In fact, in Yoshida's scenario, the entirety of the late radiation until
complete evaporation is maximally entangled with the combined system of
early radiation and the infallen matter's external reference. This late
radiation has therefore evaporated from a maximally mixed black hole;
again implying a firewall. Indeed, Yoshida's infallen qubits are
themselves `thermalized' into a random quantum-error-correction
code.\cite{hayden2007}
}

There have been other proposals for overcoming the firewall
paradox, questioning for example, the existence of the tensor product
structure typically assumed at the horizon (a structure that rigorously
exists for Rindler horizons\cite{Michel16}). For instance, in
Refs.~\onlinecite{Hsu13a} and~\onlinecite{Hsu13b} it has been noted that
momentum kicks produced by the Hawking radiation itself will lead to
the black hole evolving into a macroscopic superposition with distinct
locations. To counter this, it has been noted that when viewed in the
position basis, a firewall would occur in each branch of the
wavefunction and hence the firewall phenomenon would be unaffected by
such macroscopic superpositions.\cite{almheiri2013}

A more sustained critique of the existence of a tensor product
factorization between the degrees-of-freedom near a black hole including
ones describing the black hole as viewed from the outside and the
degrees-of-freedom very far away has been made by Raju and
colleagues.\cite{Raju1,Raju2,Raju3} They argue that such factorization,
known to be facilitated by massive gravitons, fails in the limit of
massless gravitons. Notwithstanding this, the majority of the Hawking
radiation from a large black hole is in the form of photons, for which
measurable entanglement is unproblematic. Thus, the huge entropy of
entanglement built up in the distant Hawking radiation must have
entangled partners somewhere in the vicinity of the black hole itself.
This again implies a breakdown of `no drama' in the vicinity of the
horizon and hence a firewall.\cite{Braunstein2018} In any case, the
current consensus appears to be in favor of the tensor product structure
holding at the horizon either exactly or at least to an excellent
approximation for large black holes.\cite{Alm21} 
{  There are now many other papers on the firewall paradox, for example see the review in Ref.[\onlinecite{Harlow2016}].}

{ It is also worth noting that alternative perspectives, such as the ER=EPR conjecture, 
have been proposed to reconcile the Einstein-Rosen bridge (wormholes) with quantum entanglement between black holes 
or different regions of spacetime.\cite{MaldacenaSusskind2013} Although not always discussed in the same context as firewalls, 
the ER=EPR idea underscores the intricate relationship between spacetime geometry and quantum correlations, 
reinforcing the notion that resolving the firewall paradox may demand a deeper, more unified viewpoint of quantum gravity.}

Quantum scrambling denotes the dispersion of local
quantum information into its neighborhoods and finally the entire
system. Within a scrambling time, a quantum state has a random unitary
operator applied to it; and each subsequent scrambling time would lead
to the application of a new randomly selected unitary operator. Thus, a
pure quantum state would be mapped to a random pure state, which would
be further mapped to a new random pure state with each additional
scrambling time. Indeed, the application of this effect is widely
studied in the literature on black hole dynamics. By assuming that the
radiation from a black hole is always a subsystem of a random pure
state, Page proved that the entropy of the radiation will first increase
and then decrease.\cite{page1993information} He assumed that a black
hole is a fast scrambler, without providing a concrete realization of
it. Following this, Hayden and Preskill, for the first time, explicitly
proposed that black holes obey fast random unitary transformations, and
showed that old black holes behave as information
mirrors.\cite{hayden2007} In their work, Hayden and Preskill argued that
the scrambling time of a black hole should scale as $O(\sqrt{S} ~
\text{log} S)$, where $S$ is the entropy of the black
hole.\cite{hayden2007} Note that the Page time is $O(M^3)$, while the
scrambling time is $O(M\, \log M)$, implying that for a large
(astrophysical) black hole ($M\gg 1$ in appropriate units), the latter
is far shorter than the former. 

By analyzing the spread of perturbations on the stretched horizon
\cite{price1986} of the D0-brane black hole and the ADS black hole,
Sekino and Susskind further showed that the scrambling time of a black
hole should be $\frac{1}{T_H}\,$log$\,S$ times a constant, where $T_H$
is the Hawking temperature of the black
hole.\cite{susskind2008,susskind2011} This scrambling time approximately
equals that proposed by Hayden and Preskill for black holes far from
extremality, which are the ones relevant to astrophysical black holes in
nature. To have a sense of this time scale, consider a solar-mass black
hole, its scrambling time is only about $10^{-1}$ second, while its Page
time is about $10^{72}$ seconds (which is more than $10^{54}$ times the
current age of the universe).\cite{susskind2012} Indeed, black holes are
believed by some to be the fastest scramblers in
nature for any system of comparable size.\cite{susskind2008} 

{  Although the original proofs of the black hole firewall
require a black hole older than the Page time, Almheiri {\it et al.}
speculated that the firewall phenomenon might already exist after the
very short scrambling time.\cite{almheiri2013} However, Susskind
disagreed with this conjecture, arguing that Almheiri {\it et al.} had
mistakenly equated the scrambled state with a generic
state.\cite{susskind2012} Susskind emphasized that while a generic state
represents a maximally-mixed state corresponding to infinite
temperature, a scrambled pure state remains globally pure. His argument
is based on the principle that quantum scrambling is itself a unitary
transformation, implying that the global temperature of a quantum system
is unaffected by scrambling.

This argument hinges on the global nature of the pure state.
However, a local infallen observer will only be coupled to part of this 
global interior, with the remainder effectively traced out.
In Section II, we utilize a 1+1-dimensional lattice quantum field model
to demonstrate that quantum scrambling of the lattice sites can indeed
result in an observed local temperature, thereby providing new insights
into the firewall debate.}

Thus, section II provides an explicit counterexample to
Susskind's argument that a random pure state will not be observed to
have a temperature. The key insight is that an observer coupled to a
limited number of degrees-of-freedom of a globally scrambling state will
couple only to the reduced state of the global system. Thus they can see
a well-defined finite temperature even for a globally scrambled pure
state. In order to explore the importance of the size of the
`neighorhood' to which the observer is coupled in a generic scenario and
for which the scrambling is not limited to lattice-site permutations, we
turn next to the methods of quantum information. In section III, we
begin by reviewing some basic concepts about black holes and quantum
fidelity and then in section IV, we show that any sufficiently small
neighborhood of a scrambled black hole will be infinitesimally close to
a maximally mixed quantum state. This implies that locally an infalling
observer will experience a high temperature as they pass the horizon. We
allow for arbitrary amounts of emitted radiation and a black hole which
may be initially pure or mixed. Finally, in section V we summarize our
conclusions.

\section*{II. Local temperature of a scrambled\\
lattice quantum field}

Before delving into the scrambling behavior of black holes,
it is instructive to first examine simplified scrambling in a lattice
quantum field. This will serve as a toy model of scrambling, not a
depiction of black hole scrambling itself. It is intended to illustrate
a conceptual point: that even if the global quantum state remains pure,
suitably randomized transformations can induce local thermal behavior
for an observer confined to a small region of the system. Although
deliberately simplified, the model captures the essence of how local
thermal signatures can emerge from global purity, a concept central to
our broader argument.

Consider a quantized real massless scalar field on a spatial lattice in
1+1-dimensional Minkowski spacetime, with lattice spacing $\delta$. The
Hamiltonian takes the form (see Appendix A)
\begin{equation}
H=\frac{1}{2\delta}\left(
\vec{\pi}^{\,T} \cdot \mathbb{1}_N \cdot \vec{\pi}
+ \vec{\phi}^{\,T} \cdot V \cdot \vec{\phi} \right),
\label{Ham}
\end{equation}
where the field $\phi_i$ at each lattice site $i$ allows us to form
the $N$-dimensional vector $\vec{\phi}= (\phi_1,\cdots,\phi_N)$ with
conjugate momentum $\vec{\pi}= (\pi_1,\cdots,\pi_N)$ satisfying
$[\phi_i,\pi_{i'}]=i\delta_{ii'}$, with natural units so that $\hbar=1$.
The interaction matrix $V$ here given by
\begin{equation}
        V =  \begin{pmatrix}
                1 & -1 & 0 & 0 & 0 & 0 \\
                -1& 2 &-1  & 0 &0  & 0 \\
                0& -1 & 2 & \ddots & 0 & 0 \\
                0& 0 & \ddots & \ddots & -1 & 0 \\
                0& 0 & 0 & -1 & 2 & -1 \\
                0& 0 &0  & 0 & -1 & 1 \\
        \end{pmatrix} .
\end{equation}

The ground state of this Hamiltonian is given by
\begin{equation}
\Psi = {\cal N} \exp\left(-\frac{1}{2} 
\vec{\phi}^{\,T} \cdot \sqrt{V} \cdot \vec{\phi} \right),
\label{grnd}
\end{equation}
see Appendix A. We consider a much reduced space of random unitary
operators which correspond simply to random permutations among
the lattice sites. This allows us to easily construct the `scrambled'
ground state by the replacement $\vec{\phi}\mapsto P \vec{\phi}$
in Eq.~(\ref{grnd}), for a random permutation operator $P$.

\begin{figure}[ht]
\vskip -0.05in
\includegraphics[width=0.3\textwidth]{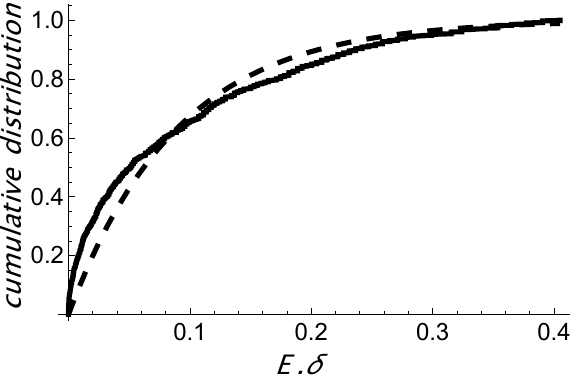}
\vskip -0.1in
\caption{Cumulative distribution for 1000 sampled energies seen by an
observer weakly coupled to the scrambled ground state at a single lattice
point (solid) on a lattice with $N=400$ sites. The fit shown is for
$k_BT.\delta=0.0893$ against the cumulative Boltzmann distribution
$1-e^{-E/k_BT}$ (dashed).}
\label{fig0}
\end{figure}

\begin{figure}[ht]
\vskip -0.1in
\includegraphics[width=0.35\textwidth]{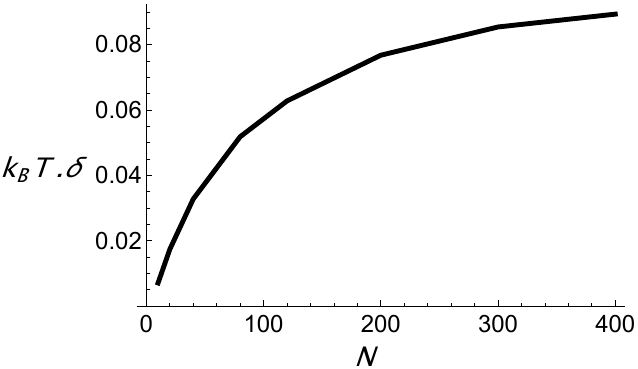}
\vskip -0.1in
\caption{Mean temperature $k_BT$ seen by an observer weakly
coupled to the scrambled ground state of a massless scalar
field at a single lattice point within a lattice of $N$ sites.
This local temperature appears to asymptote to a value
approaching the cutoff scale $O(1/\delta)$.}
\label{fig1}
\end{figure}

Consider a local observer weakly coupled to our field at a single
lattice site $i$. We can now ask what will be the expected energy seen
by such a local observer when coupled to the permuted ground state above
the actual ground state. Although each permutation yields a single
expected energy, we may consider the distribution of these energies
across random permutations. Each successive scrambling time induces a
new random permutation and consequently a new energy seen by our local
observer. By fitting this distribution to the Boltzmann distribution, we
may extract a local temperature $T$ as a function of lattice size $N$
(see the cumulative distribution fit for $N=400$ in Fig.~\ref{fig0}).
Numerical results of randomly scrambled lattices with $N\le 400$ are
shown in Fig.~\ref{fig1}.

We find that as the number of lattice sites becomes large, the
temperature seen by an observer locally coupled to our scrambled scalar
field approaches the cutoff scale $O(1/\delta)$; or within an
order-of-magnitude of that scale. Although our simplified model
of scrambling involves only the permutation of lattice sites, rather
than full-blown random unitary operators applied to the Hilbert space of
interest, the emergence of a very high temperature due to scrambling
would be expected to continue to hold for other randomization
operations, as well as in higher spacetime dimensions.

Note that unitary scrambling seamlessly translates pure
states into other pure states. Further, a pure quantum state inherently
signifies an absence of temperature. Therefore, Susskind postulated that
quantum scrambling should be ineffectual in creating any substantial
rise in temperature.\cite{susskind2012} However, our toy model has shown
that an observer who is only coupled to a limited number of
degrees-of-freedom (a limited `neighborhood') of the global quantum
system will see only the reduced state of this system. Thus, a local
observer may indeed experience a temperature. 

To move beyond the specifics of the scenario studied in this section in
order to derive a generic result we turn in the next sections to the
tools of quantum information.

\section*{III. Local state encountered by an infalling particle }

First, consider a black hole that is initially in a quantum state
described by a density matrix $\rho_0$. 
We will generalize this analysis
to a generic initial state later in this section. 
We assume its Hilbert space dimension to be $N=e^{S}$, where $S=A/4$ is the
Bekenstein-Hawking entropy of the black hole and the Boltzmann constant
is set to one. Now, if the evolution of the black hole is unitary, its
state after a unitary transformation can be written as
\begin{eqnarray}
\rho_0 \rightarrow U \rho_0 U^\dagger \equiv \rho_U
\end{eqnarray}
where $U$ is a unitary operator, i.e., $U^\dagger=U^{-1}$.

Next, consider a small particle (compared to the black hole) falling
into this black hole. It passes the horizon and interacts with a small
neighborhood surrounding it within the black hole. The quantum state of
this neighborhood, which is a tiny subsystem of the black hole, may be
obtained by tracing out the other degrees-of-freedom of the black hole.
If we assume that the dimension of this subsystem to be $n$, its state
may be expressed as $\text{tr}_{\bar n}(\rho_U)$, where ${\bar n}$
represents the degrees-of-freedom of the state complementary (i.e.,
orthogonal) to degrees-of-freedom contained within this tiny subsystem.

Since we assume that a black hole is a fast scrambler, it follows that 
the particle arrives at the black hole horizon after the scrambling
time, which is believed to be quite small for any reasonably sized black
hole.\cite{susskind2012} At this stage, the state of the black hole may
be calculated as the average over all the unitary
transformations.\cite{hayden2007} Under these conditions, we will prove
that the local quantum state `observed' by the particle (i.e., with
which it directly interacts) is almost a maximally mixed state, thereby 
having an almost infinite temperature. To show this, we will calculate
the fidelity $F$ between $\text{tr}_{\bar n}(\rho_U)$ and
${\mathbb{1}_n}/{n}$ after the scrambling time. In this paper, we define
the fidelity of two states characterized by the density matrices $\rho$
and $\sigma$ as\cite{nielsen2010} $F(\rho,\sigma) =\text{tr}
\sqrt{\sqrt{\rho}\sigma\sqrt{\rho}}$. It can be shown that
$F(\rho,\sigma) = F(\sigma,\rho)$. It can also be shown that the
fidelity of two states $\rho$ and $\sigma$ satisfies the inequalities
\cite{nielsen2010}
 {
\begin{equation}
 1- \frac{1}{2} \|\rho -\sigma \|_1 \leq F(\rho, \sigma)
\leq \sqrt{1- \frac{1}{4}\|\rho -\sigma \|^2_1}  \; ,
 	\label{Fineq1}
\end{equation}
}
\noindent
where the Schatten p-norm of $A$ is defined as
$\| A \|_p = \bigl( \text{tr}(AA^\dagger)^\frac{p}{2} \bigr)^\frac{1}{p}$.
H$\ddot{\text{o}}$lder's inequality
implies \cite{braunstein2013}
$\| \rho -\sigma\|_1 \leq \| \mathbb{1} \|_2 \times \| \rho -\sigma\|_2$,
where $\mathbb{1}$ has the same dimensionality as
$\rho$ and $\sigma$. Applying this inequality to Eq.~(\ref{Fineq1}) yields
\begin{eqnarray}
	F(\rho, \sigma) \geq 1- \frac{1}{2} \| \mathbb{1} \|_2 \|\rho -\sigma \|_2 ,
	\label{Fineq3}
\end{eqnarray}
where we have ignored the upper bound of the fidelity because it will
not play any role in the following analysis.
Since we would like to study the fidelity between $\text{tr}_{\bar
n}(\rho_U)$ and ${\mathbb{1}_n}/{n}$ after scrambling, we need to
insert these two matrices into Eq.~(\ref{Fineq3}). Under the assumption
of fast scrambling, the mean fidelity is averaged over all
unitary operators of the state of the black hole. We obtain the
following relation
\begin{eqnarray}
\Bigl\langle\! F\Bigl(\text{tr}_{\bar n}(\rho_U),\frac{\mathbb{1}_n}{n} \Bigr)
\!\Bigr\rangle_U 
	\geq 1- \frac{\sqrt{n}}{2}\!\! \int_U
\!\left\| \text{tr}_{\bar n}(\rho_U) -\frac{\mathbb{1}_n}{n} \right\|_2
\!\! dU .
	\label{fidelity}
\end{eqnarray}

\section*{IV. {Quantum states} of different black holes}

One may consider a newly formed black hole in a pure quantum state. Since
the scrambling time is very short, any small amount of Hawking radiation
emitted by the black hole as it scrambles can be safely ignored. 
%
%
For this scenario, the fidelity relation in Eq.~(\ref{fidelity}) may
be simplified to (see Appendix C)
\begin{eqnarray}
	\Bigl\langle\!F\Bigl(\text{tr}_{\bar n}(\rho_U),
	\frac{\mathbb{1}_n}{n} \Bigr)\!\Bigr\rangle_U
\geq 1- \frac{n}{2\sqrt{N}}  .
	\label{Fide1}
\end{eqnarray}
Recall that for a stellar-mass black hole $N\sim \exp({10}^{80})$. This
bound determines how close the local substate encountered by an
infalling particle is to a maximally mixed state. Since a maximally
mixed quantum state corresponds to an infinitely high temperature (see
Appendix D for a detailed analysis), this means that the infalling
object will experience a very high temperature as well.

{ 
Since the average fidelity in eq.~(\ref{Fide1}) is very close to its
maximum value of $1$, the probability that a random $\rho_U$ will
yield a fidelity, $F$, that deviates from $1$ by more than some
amount $\delta F_{\text{deviation}}$ is
\begin{equation}
        \text{prob}\bigl(F \le 1 - \delta F_{\text{deviation}}\bigr)
        \le \frac{n}{\delta F_{\text{deviation}}\,  2\sqrt{N}},
        \label{BHFubound}
\end{equation}
for $n \le \sqrt{N}$ (see the Appendix F and G). As $N$ is huge for
black holes, the probability for even a tiny deviation becomes
negligible. For example, for a stellar-mass black hole with $\log_2 n
\le (1-10^{-10}) \log_2 \sqrt{N}$ (i.e., for a local state with just
below one-half the total number of qubits of the original black hole) 
and taking $\delta F_{\text{deviation}}=e^{-10^{60}}$, we find
\begin{equation}
\text{prob}\bigl(F \le 1 - e^{-10^{60}} \bigr)
        \le e^{-10^{69}}.
\label{pbound}
\end{equation}
Thus, the average fidelity observed here is typical, indicating that
the quantum state we obtain should apply to the local states of
virtually every black hole with almost certainty.
}

To have some idea about the exact value of this temperature, we now try
to write out the expression of this quantum state. The Fidelity
$F(\rho,\sigma)$ of a pair of states $\rho$ and $\sigma$ is the maximum
overlap over all purifications $|\psi_\rho\rangle$ and $|\psi_\sigma\rangle$,
respectively, of these states, i.e.,
\begin{eqnarray}
	F\equiv F(\rho,\sigma) = \max_{\psi_\rho,\psi_\sigma}\langle
\psi_\rho | \psi_\sigma \rangle .
\label{Fid}
\end{eqnarray}
Consequently, there exist purifications $|\psi_\rho\rangle$ and
$|\psi_\sigma\rangle$ which satisfy\cite{braunstein2007}
\begin{eqnarray}
	| \psi_{\rho} \rangle
= F |\psi_\sigma \rangle + \sqrt{1-F^2}\, |\psi_\sigma^\perp \rangle,
	\label{tramix}
\end{eqnarray}
where $|\psi_\sigma^\perp \rangle$ is some quantum state orthogonal to
$|\psi_\sigma \rangle$. Taking the partial trace of this pure-state
representation yields
\begin{eqnarray}
\rho = F^2 \sigma + O\bigl(\sqrt{1-F^2}\bigr), \qquad\text{for~~} 1-F^2\ll 1.
\end{eqnarray}

Applying this result to Eq.~(\ref{Fide1}), we find that typically
\begin{eqnarray}
        \text{tr}_{\bar n}(\rho_U) &=& (1-\epsilon) \frac{\mathbb{1}_n}{n}
+ O (\sqrt{\epsilon}), \nonumber \\
&&\quad \text{for~~} \epsilon\equiv 1-F^2 \le \frac{n}{\sqrt{N}}\ll 1.
        \label{Qstate}
\end{eqnarray}
Thus, for example, for a stellar-mass black hole this reduced state of
the local neighborhood that our particle interacts with, will be in a
quantum state exceedingly close to a completely mixed state --- a state
with infinite temperature --- provided only that $n\ll \sqrt{N}$.
(see Appendix D for an analysis for how one can estimate the temperature
more precisely if we knew the Hamiltonian describing the black hole system.)

Now we consider the above black hole that has radiated a non-negligible
amount of itself away. 
If we assume the dimension of the radiation is $R$, then the dimension
of the remaining black hole will be $N_B=N/R$. The quantum state of
the remaining black hole may be represented by $\text{tr}_R (\rho_U)$.
For this scenario, the fidelity relations in Eq.~(\ref{fidelity}) equals (see Appendix E)
\begin{eqnarray}
	 F \geq 1 - \frac{n}{2\sqrt{N}} .
	\label{fidelity4}
\end{eqnarray}
Note that Eq.~(\ref{fidelity4}) is identical to Eq.~(\ref{Fide1}). So
again the quantum state of any local neighborhood encountered by an
infalling object typically has the form
\begin{eqnarray}
	\text{tr}_{\bar n} (U_B \text{tr}_R(\rho_U) {U_B}^\dagger )
= (1-\epsilon) \frac{\mathbb{1}_n}{n} + O (\sqrt{\epsilon}),
	\label{Qstate2}
\end{eqnarray}
where $0 \leq \epsilon \le \frac{n}{\sqrt{N}}\ll 1$. Therefore,
infalling objects will observe a very high temperature (see Appendix D
for how to estimate the temperature). We emphasize that this result does
not require the black hole to radiate half of itself away, which means
that this phenomenon may occur much earlier than the Page time.


In the above analysis we have assumed that the initial state of the
black hole is a pure state. However, it may seem natural that the
quantum state of an astrophysical black hole does not begin as a pure
state. Therefore, now we consider a black hole that begins
from a generic quantum state $\rho_0$, with Hilbert space dimension $N$.
If it is a newly formed black hole with negligible radiations, the
fidelity relation in Eq.~(\ref{fidelity}) will become (see Appendix F)
\begin{eqnarray}
	F \geq 1- \frac{n}{2N} \sqrt{N \operatorname{tr}
(\rho_0^2 ) - 1} 
\label{GresultR1}
\end{eqnarray}
However,
identical bounds are found when radiation is allowed for (see Appendix G).
Therefore Eq.~(\ref{GresultR1}) represents the generic result, with
$N$ the dimensionality of the original black hole, or equivalently
the product of the dimensionality of the current state of the black hole
and that of the radiation.

Since $1/N\le \text{tr}(\rho_0^2 ) < 1$ for any impure quantum state,
the lower bound to the fidelity for a black hole originating from a
non-pure state is larger than that for the pure-state scenarios studied
above. Consequently, the reduced state of a
neighborhood within an initially impure black hole will be even closer
to being a maximally mixed state than for an initially pure state black
hole.

Again, the local quantum state with which an infalling object interacts
may be written as typically given by
\begin{eqnarray}
	\text{tr}_{\bar{n}} (U \rho_0 U^\dagger)
= (1-\epsilon) \frac{\mathbb{1}_n}{n} + O (\sqrt{\epsilon}),
	\label{Qstate3}
\end{eqnarray}
where $0 \leq \epsilon \le \frac{n}{N}\sqrt{N \operatorname{tr}
(\rho_0^2) -1}\le n/\sqrt{N}\ll 1$. Therefore an infalling observer will
typically experience even higher temperatures (see Appendix D) as
impurity of the quantum state of the initial black hole is increased.

{ 
Similarly to eq.~(\ref{BHFubound}) we find
\begin{equation}
	\text{prob}\bigl(F \le 1 - \delta F_{\text{deviation}}\bigr)
	\le \frac{n \sqrt{N \operatorname{tr}
			(\rho_0^2 ) -1}}{\delta F_{\text{deviation}}\, 2N},
	\label{BHFubound2}
\end{equation}
for $n \le \sqrt{N}$ (see the Appendix F and G). Again, since
the Hilbert space dimensionality of black holes is so large the
probability for even a vanishingly small deviation from the average
behavior is itself vanishingly small, comparable to eq.~(\ref{pbound}).
}

\section*{V. Discussion}

Black holes are conjectured to be fast scramblers; possibly even the
fastest scramblers in the universe.\cite{hayden2007} The scrambling
process itself is conceived of as a random unitary operation on the
black hole interior Hilbert space.\cite{page1993information,hayden2007}
For a black hole of mass $M$, across its entire lifetime, during any
time interval of duration $O(M\log M)$, a random unitary will operate on
this subspace. This time interval is called the scrambling time.

For simplicity, let us suppose the black hole interior is initially pure
and let us ignore any evaporation process or new material being added to
the black hole. The question arises as to whether the behavior of the
quantum state of the black hole should be treated as an ensemble average
over the random unitary operators\cite{almheiri2013}, or as a single (though
randomly selected) pure state.\cite{susskind2012} In the former case,
the reduced state of a sufficiently small subsystem would appear to be
the generic maximally mixed state corresponding to an infinite
temperature.\cite{almheiri2013} In the latter case, it has been argued
that a random pure state is still pure and therefore has zero associated
temperature.\cite{susskind2012}

We have addressed this controversy head on in section II. There we study
an explicit model of a 1+1-dimensional quantum field on a lattice
undergoing random lattice site permutations (a highly restricted class
of random unitary operations on the Hilbert space of the quantum field).
We compute the energy of an observer weakly coupled to a single lattice
site for a quantum field initially in the ground (vacuum) state. Each
random unitary operator yields a distinct energy above the ground state
for this observer. After each additional scrambling time, a new random
unitary operator will cause our local observer to experience a new local
energy. As the total number of scrambled lattice sites increases, the
distribution of energies experienced by our local observer is found to
be well approximated by a Boltzmann distribution. At each point in time
the global quantum state is pure, nevertheless, local behavior is
correctly described by the ensemble statistics of the random unitary
operators (in this case permutations of lattice sites). The specific
temperature calculated in this analysis as seen by our local observer is
found to approach the natural cutoff scale of that model.

If the cutoff scale were Planckian, our toy model would generate a
scrambling temperature approaching the Planck scale. This is very close
to the temperatures found at roughly one Planck time in the Big Bang,
which like a black hole is also reputed to have a singularity at its
origin. Of course, these similar temperature scales may be just a
coincidence caused by our choice of random operations and the underlying
Hamiltonian, or it may suggest a previously unforeseen connection to the
creation of our universe and black hole physics. In the latter case, one
might attempt to model the earliest stages of the Big Bang using
scrambling dynamics. 

In any case, to overcome the model-dependent limitations associated with
relying on any specific dynamics, we consider a  more
general, information theoretic approach in section IV (with the methods
used given in section III). This allows us to take into account both the
inclusion of black hole evaporation but also an initial black hole state
which may be anywhere from completely pure to maximally mixed. We find
that if an observer is coupled to a sufficiently small
`neighborhood' of the entire interior Hilbert space they will experience
a maximally mixed state to a close approximation and therefore a near
infinite temperature. We find that a neighborhood is sufficiently small
to achieve this high-temperature behavior provided only that its
Hilbert space dimensionality, $n$, satisfies
\begin{equation}
n \le \varepsilon \sqrt{N},
\label{key}
\end{equation}
where $N$ is the dimensionality of the {\it newly formed\/} pure state
black hole. (A looser bound is found when the initial back hole is
impure which only strengthens our discussion below, see section IV for
details.) The prefactor satisfies $\varepsilon\ll 1$; for simplicity we
take $\varepsilon =2^{-10}\simeq 10^{-3}$, though using values of
$2^{-100}$ or $2^{-1000}$ makes only trivial changes to our discussion
below.

Let us take a step back and consider the scenario where an observer has
jumped into a newly formed black hole. In such a scenario, we may
neglect any Hawking radiation and consider solely the Hilbert space of
the black hole interior. The Hilbert space dimensionality of any
physically accessible degrees-of-freedom of the interior is assumed to
be given by the `central dogma' of black hole physics as $\exp(S)$ for a
black hole with Bekenstein-Hawking entropy $S$. Applying random
unitaries to this Hilbert space then implies our result of
Eq.~(\ref{key}), without needing to explicitly assume how the interior
Hilbert space is connected to the external universe, say within a tensor
product structure. Thus, if fast scrambling means anything at all on
this internal Hilbert space, Eq.~(\ref{key}) follows. 

Note further that the simplicity of Eq.~(\ref{key}) belies
the counter-intuitive nature of our results, which are technically
summarized for initially pure-state black holes in Eq.~(\ref{Fide1}) and
more generally for initially mixed-state black holes in
Eq.~(\ref{GresultR1}) gives the lower bound to the
fidelity of the fast-scrambled quantum state of a black hole as compared
to the maximally-mixed (infinite-temperature) state. For our purposes,
this bound is especially noteworthy. The counter-intuitive result is
that the larger the value of $N$ -- or in other words, the larger the
black hole -- the tighter this bound becomes. For astrophysical black
holes the fidelity is negligibly close to unity as a consequence. This
in turn implies that the infalling observer will experience an
enormously large temperature -- in effect, a firewall. This result is
very contrary to expected thinking about black holes, which is that the
larger they are, the more classical they should appear. 

In particular, rather than quantifying the size of the neighborhood in
terms of its dimensionality, it is more physically intuitive to quantify
it in terms of qubits. Note that we are not saying that any part of the
black hole is actually made up of two-level systems, only that the {\it
number\/} of two-level systems that could be supported by its
dimensionality is a more familiar quantity --- analogous to entropy the
number of qubits is additive. So the total number of qubits within the
black hole Hilbert space at any time is the sum of the number of qubits
within the selected neighborhood and the number still within the black
hole, but not within this neighborhood.

Within the language of qubits, Eq.~(\ref{key}) states that a
neighborhood of the black hole interior is sufficiently small to
correspond to an almost infinite-temperature state provided only that
\begin{equation}
\#_{\text{neighborhood}} \le \frac{1}{2}\,
\#_{\text{newly-formed-black-hole}} -10,
\label{qkey}
\end{equation}
where we denote the number of qubits within subsystem $A$ by the
hashtag $\#_A$. Replacing $\varepsilon$ by much tighter values mentioned
above only replaces the $-10$ by $-100$ or $-1000$, which as noted
are utterly trivial in comparison to values of 
$\#_{\text{newly-formed-black-hole}} \simeq 10^{80}$ for a stellar mass
black hole.

Consider an initially pure-state black hole at least as old as its
scrambling time (not very different in order of magnitude than the
light-travel time across a Schwarzschild radius' distance in flat
spacetime). Eq.~(\ref{qkey}) tells us that any body which simultaneously
couples to less than one-half of the total number of qubits in the
original black hole will experience an extremely high temperature as
soon as it passes the horizon. Thus scrambling alone places enormous
constraints on the survival of the equivalence principle's claimed
prediction of `no drama' for infalling bodies. As the black hole
evaporates this constraint becomes even more difficult to attain.

Consider an initially pure-state black hole that has partially evaporated
leaving a fraction, $f$, of the qubits in the current state of the black hole,
which therefore has
\begin{equation}
\#_{\text{BH}}=f \, \#_{\text{newly-formed-black-hole}},
\end{equation}
qubits remaining (or equivalently, it's area has shrunk to this fraction
of its original size). Then Eq.~(\ref{qkey}) becomes
\begin{equation}
\#_{\text{neighborhood}} \le \frac{1}{2f}\,
\#_{\text{BH}} -10.
\label{qkey2}
\end{equation}
Thus, as $f\rightarrow {\frac{1}{2}}^+$, (i.e., as
the evaporation approaches the Page time), any infalling body must
simultaneously couple to virtually the entire black hole interior to
have any hope of experiencing no drama has it passes the horizon. From
the Page time onwards, even this is not sufficient and we recover the
usual firewall result, though without the need for any decoding or 
complexity assumption.

In order to preserve the claim of `no drama' for scrambling black holes
even prior to the onset of a full blown firewall, any infalling body
must couple to virtually the entire interior Hilbert space of the black
hole. Further, it must do so in a uniform manner without random phases
appearing in the coupling, whatever the direction of the infalling body.
We can envisage only one scenario where this is possible: that the
quantum state of the black hole interior is actually described by a
Bose-Einstein condensate, so that all interior degrees-of-freedom
correspond to excitations of a single Bose-Einstein condensate mode.
However, we note, that even with this radical assumption, the onset of a
firewall at the Page time will occur regardless of the size of the local
neighborhood of the infalling observer.

If the black hole's quantum state is initially partially mixed, the
constraints on satisfying the equivalence principle become even more
extreme. It is convenient to define the log-purity of the {\it
initial\/} black hole as \begin{equation} \ell\equiv
-\log_2({\text{tr}}(\rho_0^2) -1/N)\ge 0, \end{equation} where $\rho_0$
is the initial black hole's density matrix and $N$ its dimensionality.
Black hole scrambling therefore implies that a fully developed firewall
will be present once the black hole's area has shrunk to the fraction
\begin{equation}
f=\frac{1}{2-\ell/\#_{\text{BH}}} \ge \frac{1}{2},
\end{equation}
of its original size. In other words, for a black hole with an initially
partially mixed quantum state, the firewall becomes fully developed
prior to the Page time regardless of the size of the local
neighborhood of the infalling observer.

As noted, the only caveats we can identify to our analysis are either
that black holes are described by a Hamiltonian that is totally
degenerate with regard to all their interior degrees-of-freedom, wherein
the notion of temperature becomes a meaningless concept, or that the
quantum state of a black hole interior is described by a Bose-Einstein
condensate. Absent these caveats we may interpret our results as proving
a new black hole paradox demonstrating the incompatibility between
the assumptions of fast scrambling, the equivalence principle's ``no
drama'' at the horizon, and the locality of an infalling observer.
Unlike the usual firewall result which holds only for old black holes,
our paradox applies to every astrophysical black hole in the universe,
which has the implication that either the astrophysical
objects observed by LIGO and the EHT are not really black holes but
something very similar (and strange), or that there is something wrong
with one of the assumptions we make about black hole physics. 
Interestingly, preliminary arguments suggest that galaxy quenching may 
be explained by a much more violent neighborhood in the vicinity of
black holes\cite{Verma2025} than can be accounted for by naive expectations
of the equivalence principle with its ideal of no drama. The conjecture
of fast scrambling within black holes may therefore be the missing
link between theoretical and astrophysical black holes.

\section*{Appendix A}

\subsubsection{Multivariate Gaussian integral}

Before calculating the energy difference between the ground state and
the scrambled ground state, we first review some results about the
multivariate Gaussian integral that we will use later.

For single variable case, we know that
\begin{eqnarray}
	\int_{-\infty}^{\infty}  e^{-a(x+b)^2}\,dx= \sqrt{\frac{\pi}{a}}.
\end{eqnarray}
Then, the multivariate Gaussian integral with linear term may be calculated as
\begin{eqnarray}
	&&\int e^{-\frac{1}{2}\vec{x}^\mathsf{T} A \vec{x}+\vec{J}^\mathsf{T} \vec{x}} d^n x  \nonumber \\
	&=& \int e^{-\frac{1}{2}[(\vec{x}-A^{-1}\vec{J})^\mathsf{T} A (\vec{x}-A^{-1}\vec{J}) -\vec{J}^\mathsf{T} A^{-1} A A^{-1} \vec{J} ]} d^n x  \nonumber \\
	&=& e^{\frac{1}{2}\vec{J}^\mathsf{T} A^{-1} \vec{J}} \int e^{-\frac{1}{2}(\vec{x}-A^{-1}\vec{J})^\mathsf{T} A (\vec{x}-A^{-1}\vec{J})} d^n x .
	\label{MultiGaus1}
\end{eqnarray}
To further simplify Eq.~(\ref{MultiGaus1}), we make the coordinate transformation $y=O \vec{x} - OA^{-1}\vec{J}$ with $O A O^\mathsf{T} $ diagonalize the matrix $A$ into $A_\text{D}$. Thus, we have
\begin{eqnarray}
	\int e^{-\frac{1}{2}\vec{x}^\mathsf{T} A \vec{x}+\vec{J}^\mathsf{T} \vec{x}} d^n x 	&=& e^{\frac{1}{2}\vec{J}^\mathsf{T} A^{-1} \vec{J} } \int e^{-\frac{1}{2}\vec{y}^\mathsf{T} A_\text{D} \vec{y}} d^n y  \nonumber \\
	&=& e^{\frac{1}{2}\vec{J}^\mathsf{T}A^{-1}\vec{J}} \sqrt{ \frac{(2\pi)^n}{\det{A}} } ,
	\label{MultiGaus2}
\end{eqnarray}
where the diagonal matrix allows us to treat this integral as multiple single-variable Gaussian integral times each other.

Then we try to derive another useful result about the multivariate Gaussian integral.
\begin{eqnarray}
	&&\!\!\! \int x_i x_j B_{ij} e^{-\frac{1}{2}\vec{x}^\mathsf{T} A \vec{x}+\vec{J}^\mathsf{T} \vec{x}} d^n x 	\nonumber \\
	&=&\!\!\! \int  B_{ij} \frac{\partial}{\partial J_i} \frac{\partial}{\partial J_j} e^{-\frac{1}{2}\vec{x}^\mathsf{T} A \vec{x}+\vec{J}^\mathsf{T} \vec{x}}  d^n x  \nonumber \\
	&=&\!\!\! B_{ij} \frac{\partial}{\partial J_i} \frac{\partial}{\partial J_j} e^{\frac{1}{2}\vec{J}^\mathsf{T}A^{-1}\vec{J}} \sqrt{ \frac{(2\pi)^n}{\det{A}} } \nonumber \\
	&=&\!\!\! B_{ij} \frac{1}{2} \frac{\partial}{\partial J_i} \left[ \left( (A^{-1})_{jk} J_k \!+\! J^\mathsf{T}_k (A^{-1})_{kj} \right)  e^{\frac{1}{2}\vec{J}^\mathsf{T}A^{-1}\vec{J}} \right] \!\! \sqrt{ \frac{(2\pi)^n}{\det{A}} } \nonumber \\
	&=&\!\!\! B_{ij} \frac{\partial}{\partial J_i} \left[ (A^{-1})_{jk} J_k  e^{\frac{1}{2}\vec{J}^\mathsf{T}A^{-1}\vec{J}} \right]  \sqrt{ \frac{(2\pi)^n}{\det{A}} } \nonumber \\
	&=&\!\!\! B_{ij} \left[ (A^{-1})_{ji} + (A^{-1})_{jk} J_k (A^{-1})_{il} J_l \right] e^{\frac{1}{2}\vec{J}^\mathsf{T}A^{-1}\vec{J}} \sqrt{ \frac{(2\pi)^n}{\det{A}} }  . \nonumber \\
	\label{MultiGaus3}
\end{eqnarray}
If we assume $\vec{J}=0$ in Eq.~(\ref{MultiGaus3}), then we will obtain
\begin{eqnarray}
	\int \vec{x}^\mathsf{T} {B} \vec{x} e^{-\frac{1}{2}\vec{x}^\mathsf{T} A \vec{x}} d^n x = \sqrt{ \frac{(2\pi)^n}{\det{A}} }  \text{tr}({B}A^{-1}) .
	\label{MultiGaus4}
\end{eqnarray}

\subsubsection{Massless scalar field based on the lattice representation }

The Lagrangian of the massless scalar field in a $(n+1)$-dimensional Minkowski space can be written as
\begin{eqnarray}
	L &=& \int \mathrm{d}^{n}x\, \mathrm{d}t\, \mathcal{L} \nonumber \\
	&=& \int \mathrm{d}^{n}x\, \mathrm{d}t \left( \frac{1}{2}\eta^{\mu\nu}\partial_\mu\phi\partial_\nu\phi \right) \nonumber \\
	&=& \frac{1}{2} \int \left( \partial_t \phi\partial_t \phi - \sum_{i=1}^n \partial_{x^{(A)}} \phi\partial_{x^{(A)}} \phi \right) \mathrm{d}^nx\, \mathrm{d}t , 
\end{eqnarray}
where $\eta^{\mu\nu}$ is the Minkowski metric.
If we define the momentum as $\pi = \frac{\partial \mathcal{L}}{\partial (\partial_t \phi)} = \partial_t \phi=\dot{\phi}$, the Hamiltonian of the massless scalar field at a given time equals
\begin{eqnarray}
	H &=& \int \left[ \dot{\phi} \pi - \mathcal{L}  \right] \mathrm{d}^nx  \nonumber \\
	&=& \frac{1}{2} \int \left( \pi^2 + \sum_{A=1}^n  (\partial_{x^{(A)}} \phi)^2 \right) \mathrm{d}^nx. 
\end{eqnarray}
{ 
Where we now use $A$ to label the spatial dimension. Thus, a general
spatial coordinate is given by $x=(x^{(A)})=(x^{(1)},\ldots,x^{(n)})$.

Let us now consider a hypercubical lattice to discretize the
$n$-dimensional spatial hypersurface. This yields a multi-index $i\ldots
j$ to label any lattice point. The $A^{\text{th}}$ coordinate of the
$i^{\text{th}}$ lattice point (which lies along the $A^{\text{th}}$
axis) is simply given by $x^{(A)}_i$. Thus, the location of a labeled
lattice point $i\ldots j$ is given by $x_{i\ldots
j}=(x^{(1)}_i,\ldots,x^{(n)}_j)$.

For simplicity, we assume that the lattice spacing along each spatial
direction is equal, so that $\delta \equiv x^{(1)}_{i+1} -
x^{(1)}_i=\cdots =x^{(n)}_{j+1} - x^{(n)}_j$ for arbitrary $i,\ldots,j$.
So an elementary spatial hypercube $\Delta_{i\ldots j}$ has one extreme
corner at $x_{i\ldots j}=(x^{(1)}_i,\ldots,x^{(n)}_j)$ and the opposite
corner at $x_{i+1\ldots j+1}=(x^{(1)}_{i+1},\ldots,x^{(n)}_{j+1})$, with
all other corners as expected, so that the volume of the hypercube
is given by
}
\begin{eqnarray}
{  \int_{ \Delta_{i\ldots j} } \!\!\! d^n x
= \int_{x^{(1)}_i}^{x^{(1)}_{i+1}} \cdots \int_{x^{(n)}_j}^{x^{(n)}_{j+1}}
\! d^n x
= \delta^n } 
\end{eqnarray}

{  We now define the discretized scalar field at the
lattice $x_{i\ldots j}$ as } 
\begin{eqnarray}
	{  \phi_{i\ldots j} } 
	&\equiv& {  \delta^{-a}  \int_{ \Delta_{i\ldots j} }
\!\!\!\phi(x_{i\ldots j})\, d^n x} \nonumber \\
	&\simeq& {  \delta^{n-a} \phi(x_{i\ldots j}) ,
\;\; \text{for} \;\; x_{i\ldots j} \in \Delta_{i\ldots j} \; . }
	\label{phi1} 
\end{eqnarray}
Similarly, the momentum may be defined as 
\begin{eqnarray}
	{  \pi_{i\ldots j}} &\equiv& 
{  \delta^{-b} \int_{\Delta_{i\ldots j}} \!\!\!\pi(x_{i\ldots j})\,
\mathrm{d}^nx } \nonumber \\
	&\simeq&{  \delta^{n-b} \pi(x_{i\ldots j}) ,
\;\; \text{for} \;\; x_{i\ldots j} \in \Delta_{i\ldots j} . }
	\label{pi1}
\end{eqnarray}
{  Here we have introduced the parameters $a$ and $b$ to ensure
both a simple canonical commutation relation and a simple Hamiltonian for
the discrete variables as we shall now see. }

With these constructions, the canonical commutation relation may be calculated as
\begin{eqnarray}
	&&\left[\int_{\Delta_{i\ldots j}} \phi(x_{i\ldots j}) \mathrm{d}^nx ,
\int_{\Delta_{i'\ldots j'}} \pi({x}_{i'\ldots j'}) \mathrm{d}^nx \right]
\nonumber \\
	&=&  \delta^{a+b} [\phi_{i\ldots j}, \pi_{i'\ldots j'}].
	\label{delta1}
\end{eqnarray}
On the other hand, the canonical commutation relation also equals
\begin{eqnarray}
	&&	\left[\int_{\Delta_{i\ldots j}} \phi(x_{i\ldots j})
\,\mathrm{d}^nx ,
\int_{\Delta_{i'\ldots j'}} \pi({x}_{i'\ldots j'})\, \mathrm{d}^nx \right]
\nonumber \\
	&=& \int_{\Delta_{i\ldots j}} \int_{\Delta_{i'\ldots j'}}
\mathrm{d}^nx \,\mathrm{d}^nx \,[\phi(x_{i\ldots j}), \pi(x_{i'\ldots j'})]
\nonumber \\
	&=& \int_{\Delta_{i\ldots j}} \int_{\Delta_{i'\ldots j'}}
\mathrm{d}^nx \,\mathrm{d}^nx \, i \, \delta (x_{i\ldots j}-x_{i'\ldots j'})
\nonumber \\
	&=& i \, \delta_{ii'} \cdots \delta_{ij'} \delta^n .
	\label{delta2}
\end{eqnarray}
Comparing Eqs.(\ref{delta1}) and (\ref{delta2}) yields
\begin{eqnarray}
[\phi_{i\cdots j}, \pi_{i'\cdots j'}] = i \, \delta_{ii'} \cdots \delta_{ij'}
\end{eqnarray}
providing $a+b =n$.
Inserting $b =n-a$ into Eqs.~(\ref{phi1}) and (\ref{pi1}) yields
\begin{eqnarray}
		{  \phi_{i \cdots j} = \delta^{n-a} \phi(x_{i\cdots j})  \;\;\;\; \text{and} \;\;\;\; \pi_{i \cdots j}= \delta^{a}	\pi(x_{i\cdots j})  . }
\end{eqnarray}

With the above results, the Hamiltonian on the lattice may be written as 
\begin{eqnarray}
	H \!\!\!&=&\!\!\! \frac{1}{2} \int \left( \pi^2 +  \sum_{A=1}^{n} (\partial_{x^{(A)}} \phi)^2 \right) \mathrm{d}^nx \nonumber \\
	&=&\!\!\! \frac{1}{2} {  \delta^n }
\sum_{i\ldots j} \left( \! \delta^{-2a} \pi_{i\ldots j}^2 +
  \frac{\delta^{2(a-n)}}{\delta^2} (\phi_{i+1 \ldots j} \!-\!
\phi_{i\ldots j})^2 +\cdots \! \right) .\nonumber \\
	\label{Hamil}
\end{eqnarray}
We choose $a$ to ensure that each term in Eq.~(\ref{Hamil}) should have
the same power of $\delta$, so we have
$-2a=2a-2n-2\Rightarrow a= \frac{n+1}{2}, b= \frac{n-1}{2}$.
Therefore, the Hamiltonian becomes
\begin{eqnarray}
	H = \frac{1}{2\delta} \sum_{i\cdots j} \left( \pi_{i\cdots j}^2 + (\phi_{i+1 \cdots j} \!-\! \phi_{i\cdots j})^2 + \cdots \right) .
	\label{Hami2}
\end{eqnarray}

In this work, we consider $n=1$ as an example, and the study of higher
dimensions should be similar. 
We then have
\begin{eqnarray}
	H &=& \frac{1}{2\delta} \sum_{i=1}^N \left( \pi_i^2 + (\phi_{i+1} \!-\! \phi_{i})^2 \right) \nonumber \\
	&=& \frac{1}{2\delta} \sum_{i=1}^N \left( \pi_i^2 +
\vec{\phi_i}^\mathsf{T} \cdot V_{ij} \cdot \vec{\phi_j} \right)  \nonumber \\
	&=& \frac{1}{2\delta} \left( \pi^\mathsf{T} \mathbb{1}_N \pi
+ \vec{\phi}^\mathsf{T} \cdot V \cdot \vec{\phi} \right) ,
	\label{Hami3}
\end{eqnarray}

where
\begin{equation}
	\vec{\phi}= (\phi_1,\cdots,\phi_N)~,
\end{equation}
\begin{equation}
	V_{ij} = \begin{pmatrix}
		1 & -1 & 0 & 0 & 0 & 0 \\
		-1& 2 &-1  & 0 &0  & 0 \\
		0& -1 & 2 & \ddots & 0 & 0 \\
		0& 0 & \ddots & \ddots & -1 & 0 \\
		0& 0 & 0 & -1 & 2 & -1 \\
		0& 0 &0  & 0 & -1 & 1 \\
	\end{pmatrix} ,
\end{equation}
and $\mathbb{1}_N$ is the $(N\times N)$ identity matrix. 

\subsubsection{Vacuum state, scrambled vacuum state and their energy}

The ground state of the above scalar field in the lattice representation is written as a Gaussian as follows
\begin{eqnarray}
	\Psi = {\cal N} e^{-\frac{1}{2}\vec{\phi}^\mathsf{T} \cdot Q \cdot \vec{\phi}} ~,
	\label{groundstate1}
\end{eqnarray}
where $Q$ is a $(N\times N)$ matrix to be determined below. 
Recall that the canonical commutation relation $ [ x_i, \pi_j ]= i \delta_{ij}$ implies $\pi_j=-i \frac{\partial}{\partial x_j}$. Therefore, Eq.~(\ref{Hami3}) can be written as
\begin{eqnarray}
	H = \frac{1}{2\delta} \left( - \delta_{ij} \frac{\partial}{\partial \phi_i} \frac{\partial}{\partial \phi_j}
+ \vec{\phi}^\mathsf{T} \cdot V \cdot \vec{\phi} \right) .
	\label{Hami4}
\end{eqnarray}
Applying this Hamiltonian operator to the ground state yields
\begin{eqnarray}
H \Psi &=& \frac{{\cal N}}{2\delta} \left( - \delta_{ij} \frac{\partial}{\partial \phi_i} \frac{\partial}{\partial \phi_j} + \vec{\phi}^\mathsf{T} \cdot V \cdot \vec{\phi} \right) e^{-\frac{1}{2}\vec{\phi}^\mathsf{T} \cdot Q \cdot \vec{\phi}} \nonumber \\
&=& \frac{{\cal N}}{2\delta} \left( Q_{ii} - \delta_{ij} Q_{ik} \phi_k Q_{jl} \phi_l + \vec{\phi}^\mathsf{T} \cdot V \cdot \vec{\phi} \right) e^{-\frac{1}{2}\vec{\phi}^\mathsf{T} \cdot Q \cdot \vec{\phi}} \nonumber \\
&=& \frac{1}{2\delta} \left( \text{tr}\,Q + \vec{\phi}^\mathsf{T} \cdot (V-Q^2) \cdot \vec{\phi} \right) \Psi  ,
\label{Hami5}
\end{eqnarray}
where a similar technique as in 
Eq.~(\ref{MultiGaus3}) has been used in going 
from the first to the second line.

Now since Eq.~(\ref{groundstate1}) is an eigenstate of $H$, it follows from Eq.~(\ref{Hami5}) 
that $Q=\sqrt{V}$.
Thus, the ground state simply becomes
\begin{eqnarray}
	\Psi = {\cal N} e^{-\frac{1}{2}\vec{\phi}^\mathsf{T} \cdot \sqrt{V} \cdot \vec{\phi}} ,
	\label{groundstate2}
\end{eqnarray} 
with the corresponding ground state energy 
\begin{eqnarray}
	E_0= \frac{1}{2\delta} \text{tr} \sqrt{V}.
	\label{energy1}
\end{eqnarray}
To normalize the ground state, we require $\int \Psi^2 d^N \phi=1$, i.e.,
\begin{eqnarray}
	1 &=& {\cal N}^2 \int e^{-\frac{1}{2}\vec{\phi}^\mathsf{T} \cdot 2\sqrt{V} \cdot \vec{\phi}} d^N \phi = {\cal N}^2 \sqrt{ \frac{(2\pi)^N}{\det{(2\sqrt{V}})} } \nonumber \\
	&=& {\cal N}^2 \sqrt{ \frac{\pi^N}{\det{\sqrt{V}}} } ,
	\label{groundstate3}
\end{eqnarray}
where we have used Eq.~(\ref{MultiGaus2}). 
This implies 
\begin{eqnarray}
	{\cal N} = \frac{(\det V)^{\frac{1}{8}}}{\pi^{\frac{N}{4}}} .
\end{eqnarray}

Next, if we randomly permute (scramble) the scalar field $\phi_i$ at the lattice points, such that 
$\vec\phi\rightarrow P\,\vec\phi$, where 
$P$ is the permutation operator, 
and repeat the above analysis, we obtain similar results. Namely, the scrambled Hamiltonian becomes
\begin{eqnarray}
	H = \frac{1}{2\delta} \left( \pi^\mathsf{T} \mathbb{1} \pi
+ \vec{\phi}^\mathsf{T} \cdot K \cdot \vec{\phi} \right) ,
	\label{Hami6}
\end{eqnarray}
where $K=P V P^\mathsf{T} $ is the new potential matrix in the randomly permuted $\vec{\phi}$.
%
The scrambled ground state is now
\begin{eqnarray}
	\Psi = \frac{(\det K)^{\frac{1}{8}}}{\pi^{\frac{N}{4}}} e^{-\frac{1}{2}\vec{\phi}^\mathsf{T} \cdot \sqrt{K} \cdot \vec{\phi}} ,
	\label{groundstate4}
\end{eqnarray} 
with the ground state energy
\begin{eqnarray}
	E'_0 &=& \text{tr} \sqrt{K} \!=\!\frac{1}{2\delta} \text{tr} \sqrt{P \sqrt{V} P^\mathsf{T} P \sqrt{V} P^\mathsf{T}} \nonumber \\
	&=& \frac{1}{2\delta} \text{tr} (P \sqrt{V} P^\mathsf{T} ) = \frac{1}{2\delta} \text{tr} (\sqrt{V} P^\mathsf{T} P) \nonumber \\
	&=& \frac{1}{2\delta} \text{tr} \sqrt{V} .
	\label{energy2}
\end{eqnarray}
which is identical to Eq.~(\ref{energy1}). 
In fact, scrambling the Hamiltonian and ground state simultaneously is akin to using a different coordinate system to describe the same physics, such that the energy does not change.
%
%
Therefore to measure the scrambling effect, we need to calculate the mean value of the original Hamiltonian Eq.~(\ref{Hami3}) in the scrambled ground state Eq.~(\ref{groundstate4}):
\begin{eqnarray}
	E &=& \langle H \rangle = \int (\Psi^\ast H \Psi) d^N \phi \nonumber \\
	&=& \frac{{\cal N}^2}{2\delta} \int \left( \pi^\mathsf{T} \mathbb{1} \pi + \vec{\phi}^\mathsf{T}  V  \vec{\phi} \right) e^{-\frac{1}{2}\vec{\phi}^\mathsf{T} 2\sqrt{K} \vec{\phi}} d^N \phi \nonumber \\
	&=& \frac{{\cal N}^2}{2\delta} \int  \left(\text{tr} \sqrt{K} + \vec{\phi}^\mathsf{T}  (V-K) \vec{\phi} \right) e^{-\frac{1}{2}\vec{\phi}^\mathsf{T} 2\sqrt{K}  \vec{\phi}} d^N \phi \nonumber \\
	&=& \frac{1}{2\delta} \left(\text{tr} \sqrt{K} + \text{tr}[(V-K)(2\sqrt{K})^{-1}] \right) \nonumber \\
	&=& \frac{1}{2\delta} \left(\text{tr} \sqrt{V} + \frac{1}{2} \text{tr}[(V-K)(\sqrt{K})^{-1}] \right)
	\label{energy3}
\end{eqnarray}
where Eq.~(\ref{MultiGaus4}) has been used in moving from the third to the four line.

Therefore, the energy difference before and after scrambling equals
\begin{eqnarray}
	\Delta E = E-E_0 = \frac{1}{4\delta} \text{tr}[(V-K)(\sqrt{K})^{-1}] \;, 
	\label{energy4}
\end{eqnarray}
where we have used Eqs.~(\ref{energy1}) and (\ref{energy3}).

{   The above calculation gives the total energy difference for the whole lattice field before and after the fast quantum scrambling. If we assume a local observer is interacting with the $i^\text{th}$ lattice,  the Hamiltonian of the lattice field may be written }
\begin{eqnarray}
	{   H_i = \frac{1}{2 \delta} ( \pi_i^2 + V_{ii} \phi_i^2 ) = \frac{1}{2 \delta} \Bigl( - \Bigl(\frac{\partial}{\partial \phi_i} \Bigr)^2 + 2 \phi_i^2 \Bigr) , }
\end{eqnarray}
{   where we have used $\pi_i = -i \frac{\partial}{\partial \phi_i}$ and we have assumed that $i \neq 1, N$ for simplicity. Therefore, the energy of this lattice field felt by the observer equals}
\begin{eqnarray}
	{  E^0_i }  &=& {   \langle H_i \rangle = \int (\Psi^\ast H_i \Psi) d^N \phi } \nonumber \\
	&=& \!\!\! {  \frac{{\cal N}^2}{2\delta} \int e^{-\frac{1}{2}\vec{\phi}^\text{T} \sqrt{V} \vec{\phi}} \Bigl( - \Bigl(\frac{\partial}{\partial \phi_i} \Bigr)^2 + 2 \phi_i^2 \Bigr) e^{-\frac{1}{2}\vec{\phi}^\text{T} \sqrt{V} \vec{\phi}} d^N \phi } \nonumber \\
	&=& \!\!\! {   \frac{{\cal N}^2}{2\delta} \int  \Bigl( (\sqrt{V})_{ii} - (\sqrt{V})_{ik} \phi_k \phi_l (\sqrt{V})_{li} }  \nonumber \\ 
	&& {   + 2 \phi^2_i \Bigr) e^{-\frac{1}{2}\vec{\phi}^\text{T} 2\sqrt{V}  \vec{\phi}} d^N \phi }  \nonumber \\
	&=& {    \frac{ (\sqrt{V})_{ii}}{2\delta} - \frac{(\sqrt{V})_{ik} ((\sqrt{V})^{-1})_{kl} (\sqrt{V})_{li}}{4\delta} +  \frac{((\sqrt{V})^{-1})_{ii}}{2\delta}  }   \nonumber \\
	&=&  {  \frac{(\sqrt{V})_{ii}}{4\delta}  + \frac{((\sqrt{V})^{-1})_{ii}}{2\delta}  } , 
\end{eqnarray}
{   where Eq.~(\ref{MultiGaus3}) is used in moving from third to the fourth line. }

{   Similarly, after the scrambling, the energy of the lattice field at the $i^\text{th}$ lattice equals }
\begin{eqnarray}
	{    E_i } &=& \!\!\! {   \langle H_i \rangle = \int (\Psi^\ast H_i \Psi) d^N \phi } \nonumber \\
	&=& \!\!\!  {   \frac{{\cal N}^2}{2\delta} \int e^{-\frac{1}{2}\vec{\phi}^\text{T} \sqrt{K} \vec{\phi}} \Bigl( - \Bigl(\frac{\partial}{\partial \phi_i} \Bigr)^2 + 2 \phi_i^2 \Bigr) e^{-\frac{1}{2}\vec{\phi}^\text{T} \sqrt{K} \vec{\phi}} d^N \phi } \nonumber \\
	&=& \!\!\! {   \frac{{\cal N}^2}{2\delta} \int  \Bigl( (\sqrt{K})_{ii} - (\sqrt{K})_{ik} \phi_k \phi_l (\sqrt{K})_{li} }  \nonumber \\ 
	&& {   + 2 \phi^2_i \Bigr) e^{-\frac{1}{2}\vec{\phi}^\text{T} 2\sqrt{K}  \vec{\phi}} d^N \phi }  \nonumber \\
	&=&\!\!\! { \frac{(\sqrt{K})_{ii}}{2\delta} - \frac{(\sqrt{K})_{ik} ((\sqrt{K})^{-1})_{kl} (\sqrt{K})_{li}}{4\delta} + \frac{((\sqrt{K})^{-1})_{ii}}{2\delta} }  \nonumber \\
	&=&\!\!\! { \frac{(\sqrt{K})_{ii}}{4\delta} + \frac{((\sqrt{K})^{-1})_{ii}}{2\delta} }  ,
\end{eqnarray}
{  It follows from Eq.~(\ref{energy4}) that we can numerically simulate the change $\Delta E_i \equiv E_i -E^0_i$ with respect to the number of contiguous sites being scrambled, see Fig.~1 in the manuscript. }

\section*{Appendix B}

Let us first review a result of the Schur-Weyl duality. Suppose we have
a Hermitian $X$, then Schur-Weyl duality implies that\cite{abeyesinghe2009}
\begin{eqnarray}
\int_{U(A)}\left(U^\dagger \otimes U^\dagger \right) X (U \otimes U) d U
= \alpha_{+} \Pi_{+}^A+\alpha_{-} \Pi_{-}^A \;,
\label{schur1}
\end{eqnarray}
where 
\begin{equation}
\alpha_{\pm}=\frac{\operatorname{tr}\left(X \Pi_{\pm}^A\right)}
{\operatorname{rank}\left(\Pi_\pm^A\right)} \;\; \text{and} \;\;
\Pi_\pm^A=\frac{1}{2}\left(\mathbb{1}_{A, A} \pm \text{SWAP}_{A, A}\right) .
\label{asist1}
\end{equation}
Here SWAP$_{A,B}$ represents the SWAP operator between the system
$A$ and $B$. 

Now suppose that $A=A_1 \otimes A_2$, which is what we will use in this
work, then we have  
\begin{eqnarray}
\text{rank} \Pi_\pm^A &=&  \text{rank} \frac{1}{2}\left(\mathbb{1}_{A, A}
 \pm \text{SWAP}_{A, A}\right) \nonumber \\ 
&=& \frac{1}{2}(d_A^2 \pm d_A) \nonumber \\ 
&=& \frac{1}{2}((d_{A_1} d_{A_2})^2 \pm d_{A_1} d_{A_2}) \; ,
\label{asist2}
\end{eqnarray}
where $d_X$ equals the dimension of the state $X$.
If we suppose $X$ equals $\text{SWAP}_{A_{2}, A_2}$, then
$\operatorname{tr}\left(X \Pi_{\pm}^A\right)$ in Eq.~(\ref{asist1})
may be calculated as
\begin{widetext}
\begin{eqnarray}
\operatorname{tr}\left(\Pi_\pm^A \, \text{SWAP}_{A_{2}, A_2}\right)
&=& \frac{1}{2} \text{tr}\left[\left(\mathbb{1}_{A, A} \pm
\text{SWAP}_{A_{1}, A_{1}} \otimes \text{SWAP}_{A_2 , A_2 }\right)
\text{SWAP}_{A_2, A_2 }\right] \nonumber \\ 
&=& \frac{1}{2}\left[\operatorname{tr}\left(\mathbb{1}_{A_1, A_1 }
\otimes \text{SWAP}_{A_2, A_2 }\right) \pm \operatorname{tr}
\left(\text{SWAP}_{A_1, A_1 } \otimes \mathbb{1}_{A_{1} , A_{2} }
\right)\right]  \nonumber \\ 
&=& \frac{1}{2}\left(d_{A_1}^2 d_{A_2} \pm d_{A_1} d_{A_2}^2 \right) .
\label{asist3}
\end{eqnarray}
Inserting Eqs.~(\ref{asist1}), (\ref{asist2}) and (\ref{asist3})
into Eq.~(\ref{schur1}) yields
\begin{eqnarray}
\int_{U(A)}\left(U^\dagger \otimes U^\dagger \right) X (U \otimes U) d U
&=& \frac{d_{A_1}^2 d_{A_2} + d_{A_1} d_{A_2}^2 }{(d_{A_1} d_{A_2})^2
+ d_{A_1} d_{A_2}} \Pi_{+}^A  +\frac{d_{A_1}^2 d_{A_2}
 - d_{A_1} d_{A_2}^2 }{(d_{A_1} d_{A_2})^2 - d_{A_1} d_{A_2}}
\Pi_{-}^A  \nonumber \\ 
&=& \frac{d_{A_1} + d_{A_2} }{d_{A_1} d_{A_2} + 1} \Pi_{+}^A
 + \frac{d_{A_1} - d_{A_2} }{d_{A_1} d_{A_2} - 1} \Pi_{-}^A \nonumber \\ 
&=& \frac{d_{A_1} + d_{A_2} }{d_A + 1} \Pi_{+}^A + \frac{d_{A_1}
- d_{A_2} }{d_A - 1} \Pi_{-}^A  \;.
\label{schur2}
\end{eqnarray}

\section*{Appendix C}

To prove that the local quantum state of the black hole `observed' by the particle (i.e., with
which it directly interacts) is almost a maximally mixed state, we will calculate
the fidelity $F$ between $\text{tr}_{\bar n}(\rho_U)$ and
${\mathbb{1}_n}/{n}$ after the scrambling time. In this paper, we define
the fidelity of two states characterized by the density matrices $\rho$
and $\sigma$ as\cite{nielsen2010} $F(\rho,\sigma) =
\text{tr}\sqrt{\sqrt{\rho}\sigma\sqrt{\rho}}$. It can be shown that
$F(\rho,\sigma) = F(\sigma,\rho)$. { Since $1-F(\rho, \sigma) \leq \frac{1}{2} \|\rho -\sigma \|_1 \leq \sqrt{1-F(\rho, \sigma)^2}$, the
fidelity of two states $\rho$ and $\sigma$ satisfies the inequalities
\cite{nielsen2010}
\begin{eqnarray}
 1- \frac{1}{2} \|\rho -\sigma \|_1 \leq F(\rho, \sigma)
\leq \sqrt{ 1- \frac{1}{4}\|\rho -\sigma \|^2_1}  \; ,
 	\label{APFineq1}
\end{eqnarray} 
} 
where the Schatten p-norm of $A$ is defined as
$\| A \|_p = \bigl( \text{tr}(AA^\dagger)^\frac{p}{2} \bigr)^\frac{1}{p}$. { Since only the lower bound of the fidelity in Eq.~(\ref{APFineq1}) plays the key role in the following analysis, we will disregard the upper bound from this point onward.}
Since H$\ddot{\text{o}}$lder's inequality
implies \cite{braunstein2013}
$\| \rho -\sigma\|_1 \leq \| \mathbb{1} \|_2 \times \| \rho -\sigma\|_2$, Eq.~(\ref{APFineq1}) yields 
\begin{eqnarray}
	F(\rho, \sigma) \geq 1- \frac{1}{2} \| \mathbb{1} \|_2 \|\rho -\sigma \|_2 .
	\label{APFineq3}
\end{eqnarray}

Since we would like to study the fidelity between $\text{tr}_{\bar
n}(\rho_U)$ and ${\mathbb{1}_n}/{n}$ after scrambling, we need to
insert these two matrices into Eq.~(\ref{APFineq3}). Under the assumption
of fast scrambling, the mean fidelity is averaged over all
unitary operators of the state of the black hole. We obtain the
following average fidelity as
\begin{eqnarray}
	\Bigl\langle F\Bigl(\text{tr}_{\bar n}(\rho_U),
\frac{\mathbb{1}_n}{n} \Bigr) \Bigr\rangle_U
	\geq 1- \frac{1}{2} \int_U \| \mathbb{1}_n \|_2 \times
\left\| \text{tr}_{\bar n}(\rho_U) -\frac{\mathbb{1}_n}{n}
\right\|_2 dU = 1- \frac{1}{2} \sqrt{n} \int_U \left\| \text{tr}_{\bar n}(\rho_U)
-\frac{\mathbb{1}_n}{n} \right\|_2 dU .
	\label{APfidelity}
\end{eqnarray}

Since the scrambling time is very short, a newly formed
black hole in a pure state, yet after its scrambling,
should have negligible radiations.
For such a black hole, the two
norm $\| \text{tr}_{\bar n}(\rho_U) -\frac{\mathbb{1}_n}{n} \|^2_2$ that
appeared in the fidelity relation Eq.~(\ref{APfidelity}) in the manuscript may be calculated as
\begin{eqnarray}
\int_U \left\| \text{tr}_{\bar{n}} (\rho_U)-\frac{\mathbb{1}_n}{n}
\right\|^2_2 d U 
&=& \int_U \operatorname{tr}_{n}\left[
\operatorname{tr}_{\bar n} (\rho_U ) \operatorname{tr}_{\bar n} (\rho_U )
-\frac{2}{n} \operatorname{tr}_{\bar n}\left(\rho_U\right)
+\frac{\mathbb{1}_n}{n^{2}}\right] d U \nonumber \\
&=& \int_U \operatorname{tr}_{n}\left[
\operatorname{tr}_{\bar n} (\rho_U )
\operatorname{tr}_{\bar n} (\rho_U ) \right] d U -\frac{1}{n} \nonumber \\
&=& \int_U \operatorname{tr}_{n,n}\left[
\operatorname{tr}_{\bar n, \bar n} (\rho_U \otimes \rho_U ) \,
\text{SWAP}_{n,n} \right] d U -\frac{1}{n} \nonumber \\
&=& \int_U \operatorname{tr} \left[ (U \otimes U)
(\rho \otimes \rho) (U^\dagger \otimes U^\dagger)
(\text{SWAP}_{n,n} \otimes \mathbb{1}_{\bar n, \bar n} ) \right] d U
-\frac{1}{n} \nonumber \\
&=& \operatorname{tr} \Biggl[ (\rho \otimes \rho)  \int_U
\left[(U^\dagger \otimes U^\dagger) \, \text{SWAP}_{n,n}  (U \otimes U) 
\right] d U \Biggr]-\frac{1}{n}, \nonumber \\
\label{result1}
\end{eqnarray}
where $\| A\|^2_2 = \text{tr}(A^\dagger A)$ is applied to the first
step, and $\text{tr}(AB) = \text{tr}( A \otimes B \; \text{SWAP}_{A,B} )$
is applied in moving from the second to the third line. Here
$\text{SWAP}_{A,B}$ is the SWAP operator between the quantum subsystems
$A$ and $B$. Applying the Schur-Weyl duality in Appendix B to Eq.~(\ref{result1}) then yields
\begin{eqnarray}
	\int_U \left\| \text{tr}_{\bar{n}} (\rho_U)-\frac{\mathbb{1}_n}{n}
	\right\|^2_2 d U  &=& \frac{1}{2} \operatorname{tr} \Biggl[ (\rho \otimes \rho) 
	\left(\frac{\frac{N}{n}+n}{N+1}(\mathbb{1}_{N,N}+ \text{SWAP}_{N,N})
	+  \frac{\frac{N}{n}-n}{N-1}(\mathbb{1}_{N,N}
	- \text{SWAP}_{N,N}) \right) \Biggr]-
	\frac{1}{n} \nonumber \\
	&=& \operatorname{tr} \Biggl[ (\rho \otimes \rho) 
	\left(\frac{\frac{N^2}{n}-n}{N^2-1} \mathbb{1}_{N,N}
	+ \frac{Nn-\frac{N}{n}}{N^2-1} \text{SWAP}_{N,N} \right) \Biggr]
	- \frac{1}{n} \nonumber \\
	&=& \frac{\frac{N^2}{n}-n}{N^2-1} (\operatorname{tr} \rho )^2
	+ \frac{Nn-\frac{N}{n}}{N^2-1} \operatorname{tr} (\rho^2 )
	- \frac{1}{n} = \frac{\frac{N^2}{n}-n}{N^2-1}
	+ \frac{Nn-\frac{N}{n}}{N^2-1} - \frac{1}{n}  \nonumber \\
	&=& \frac{Nn^2-N-n^2+1}{(N^2-1)n} = \frac{n^2-1}{(N+1)n}
	= \frac{1}{N+1} (n-\frac{1}{n}) ,
	\label{result}
\end{eqnarray}
where $(\operatorname{tr} \rho )=\operatorname{tr} (\rho^2 )=1$ is
used in the third line because the entire black hole is a pure quantum state.

Since $N \geq n \geq 1$, Eq.~(\ref{result}) may be further simplified as
\begin{eqnarray}
	\int_U \left\| \text{tr}_{\bar{n}} (\rho_U)-\frac{\mathbb{1}_n}{n}
	\right\|^2_2 d U <  \frac{n}{N} .
	\label{result2}
\end{eqnarray}
Inserting Eq.~(\ref{result2}) into Eq.~(\ref{APfidelity}), then the average fidelity relation in Eq.~(\ref{APfidelity}) may
be written 
\begin{eqnarray}
	\Bigl\langle F\Bigl(\text{tr}_{\bar n}(\rho_U),
\frac{\mathbb{1}_n}{n}\Bigr) \Bigr\rangle_U
\geq 1- \frac{n}{2\sqrt{N}} .
	\label{BlackHoleFidelity}
\end{eqnarray}

{   Since the dimensionality of black holes is immense, for a
local state with dimensionality less than half the black hole's total
interior ($n \leq \sqrt{N}$), the fidelity in
Eq.~(\ref{BlackHoleFidelity}) will be close to 1. We now demonstrate
that this high average fidelity implies the probability of a random
black hole local state having a significant deviation in fidelity from 1
is very small. 

Each random unitary $U$ will yield a distinct fidelity. The Haar measure
over $U$ therefore induces 
a probability distribution $\text{prob}(f)$,
where $f$ denotes the fidelity variable which ranges between 0 and 1.
We may now write
\begin{eqnarray}
	1- \frac{n}{2\sqrt{N}}
	&\le& \Bigl\langle F\Bigl(\text{tr}_{\bar n}(\rho_U),
          \frac{\mathbb{1}_n}{n}\Bigr)\Bigr\rangle_U
 = \int_0^1 f \,\text{prob}(f)\, df
	= \int_0^{F_{\text{bound}}} f\, \text{prob}(f)\, df
	+ \int_{F_{\text{bound}}}^{1} f\, \text{prob}(f)\, df
	\nonumber \\
	&\le& F_{\text{bound}} \int_0^{F_{\text{bound}}} 
	\text{prob}(f)\, df
	+ 1 \times \int_{F_{\text{bound}}}^{1} \text{prob}(f)\,
	df = F_{\text{bound}} \,q + 1 \times (1-q) \nonumber \\
	&=& 1 - \delta F_{\text{deviation}}\,q ,
\end{eqnarray}
where $q\equiv \text{prob}(F \le F_{\text{bound}})$
and the deviation from maximum fidelity is defined as
$\delta F_{\text{deviation}}\equiv 1-F_{\text{bound}}$. 

Given that the average fidelity in Eq.~(\ref{BlackHoleFidelity}) is
very close to 1 for our case, the probability for a random local
state to have a fidelity $F$ less than 1 by more than a deviation
$\delta F_{\text{deviation}}$ is
\begin{equation}
	\text{prob}\bigl(F\bigl(\text{tr}_{\bar n}(\rho_U),\mathbb{1}_n/n\bigr)
\le 1 - \delta F_{\text{deviation}}\bigr)
	\le \frac{n}{\delta F_{\text{deviation}}  2\sqrt{N}} .
	\label{BHFideBound1}
\end{equation}

}

\section*{Appendix D}

In statistical mechanics, a Boltzmann distribution describes the
probability distribution of a system for different microstates $i$ with
respect to the state's energy $E_i$ and temperature of the system $T$.
The distribution may be written
\begin{equation}
p_i = \frac{1}{Z} e^{-{E_i}/{(k_\text{B}T)}} ,
\label{Boltzmann}
\end{equation}
where function $Z$ is used to normalized the distribution and hence
equals the sum of $e^{-{E_i}/{(k_\text{B}T)}}$ for all different $E_i$.

The generalization of Eq.~(\ref{Boltzmann}) to a quantum mechanics may
be written
\begin{equation}
    \hat \rho = \frac{1}{Z}\exp\big(-\frac{\hat H}{k_\text{B}T}\big),
\label{BolQuan}
\end{equation}
where $\hat \rho$ is the density matrix of the quantum system, $\hat H$
is the system's Hamiltonian operator, and $\exp$ represents a
exponential function for quantum operator (or matrix). Here, the
function $Z$ is used to normalize the density matrix of the quantum
state and equals the trace of $\exp (-\hat H/(k_\text{B}T))$.

Now we prove that Eq.~(\ref{BolQuan}) is the quantum version of the
Boltzmann distribution Eq.~(\ref{Boltzmann}). Suppose the
basis states of the quantum system are labeled by $i$ and $ \{| i\rangle
\}$ is a complete basis, then Eq.~(\ref{BolQuan}) may be written
\begin{eqnarray}
 \hat \rho = \frac{1}{Z}\exp\big(-\frac{\hat H}{k_\text{B} T}\big)
= \frac{1}{Z} \sum_{i,j} |i\rangle \langle i |
\exp\big(-\frac{\hat H}{k_\text{B} T}\big) |j \rangle \langle j |
= \sum_{i,j} p_{i,j} |i \rangle \langle j | , \;\; \text{where} \;\;
p_{i,j} = \frac{1}{Z}  \langle i | \exp\big(-\frac{\hat H}
{k_\text{B} T}\big) |j \rangle .
 \label{BolQuan2}
\end{eqnarray}
Here $p_{i,j}$ is the probability to find the quantum system in the
quantum state $|i \rangle \langle j |$ which has the similar physical
interpretation as $p_i$ in the Boltzmann distribution
Eq.~(\ref{Boltzmann}).

If we choose the basis $ \{| i\rangle \}$ as a complete basis
of the energy eigenstates, the $ p_{i,j}$ in Eq.~(\ref{BolQuan2}) may be
further simplified
\begin{equation}
p_{i,j} = \frac{1}{Z}  \langle i | \exp\big(-\frac{\hat H}
{k_\text{B} T}\big) |j \rangle
= \frac{1}{Z}  \langle i | |j \rangle e^{-\frac{E_j}{k_\text{B} T}}
= \frac{1}{Z} e^{-\frac{E_j}{k_\text{B} T}} \delta_{i,j} .
\label{delta}
\end{equation}
where $E_j$ is the eigenvalues $| j \rangle$.
Inserting Eq.~(\ref{delta}) into Eq.~(\ref{BolQuan2}) yields
\begin{eqnarray}
 \hat \rho = \sum_{i,j} \frac{1}{Z} e^{-\frac{E_j}{k_\text{B} T}}
\delta_{i,j} |i \rangle \langle j |
= \sum_{i} \frac{1}{Z} e^{-\frac{E_i}{k_\text{B} T}} |i \rangle \langle i | ,
 \label{BolQuan3}
\end{eqnarray}
where $Z = \sum_{i=1}^N e^{\frac{- E_i}{k_\text{B} T}}$ and $N$ is total
number of the basis eigenvectors. The density matrix is diagonal in this
basis and each entries gives the probability for the quantum system with
a specific energy. 

If we assume the temperature of the system to be high
($\frac{E_j}{k_\text{B} T}$ is small), then Eq.~(\ref{BolQuan3}) may be
approximated by 
\begin{eqnarray}
 \hat \rho &=& \sum_{i} \frac{1}{\sum_{j=1}^N e^{\frac{- E_j}{k T}}} e^{-\frac{E_i}{k_\text{B} T}} |i \rangle \langle i |  
 \approx  \sum_{i} \frac{1-\frac{E_i}{k_\text{B} T} + O(\frac{1}{k_\text{B} T})}{\sum_{n=1}^N (1-\frac{E_j}{k_\text{B} T} + O(\frac{1}{k_\text{B} T}) ) } |i \rangle \langle i |  \nonumber \\
 &=& \sum_{i} \frac{1-\frac{E_i}{k_\text{B} T} + O(\frac{1}{k_\text{B} T})}{N - \sum_{j=1}^N \frac{E_j}{k_\text{B} T} + O(\frac{1}{k_\text{B} T}) } |i \rangle \langle i |  
 = \sum_{i} \frac{1}{N} \frac{1-\frac{E_i}{k_\text{B} T} + O(\frac{1}{k_\text{B} T})}{1 - \frac{\langle E \rangle}{k_\text{B} T} + O(\frac{1}{k_\text{B} T}) } |i \rangle \langle i |  \nonumber \\
 &=& \sum_{i} \frac{1}{N} \Bigl( 1 + \frac{\langle E \rangle-E_i}{k_\text{B} T} + O(\frac{1}{k_\text{B} T} \Bigr) |i \rangle \langle i |  ,
 \label{BolQuan4}
\end{eqnarray}
where $\langle E \rangle= (\sum_{j=1}^N E_j)/N$. When the temperature is
infinity high, the quantum state of the system Eq.~(\ref{BolQuan4})
reduces to $\frac{1}{N} \mathbb{1}_N$. Thus a maximally mixed quantum
state corresponds to an infinity high temperature for the quantum
system. { The caveat to the above analysis is a scenario where the system Hamiltonian is totally degenerate. However, this is a scenario where energy and temperature have no meaning.}

\section*{Appendix E}

For a black hole that begins from a pure state with non-negligible radiations, we have the following analysis.
If we still assume the dimension of initial state of the black hole to be $N$ and the dimension of the radiation to be $R$, then the dimension
of the remaining black hole will be $N_B=N/R$. The quantum state of
the remaining black hole may be represented by $\text{tr}_R (\rho_U)$.
The key step in our calculation of the 2-norm is to take an average over
all possible unitary operators, since black holes are fast scramblers.
Now, with part of the initial state radiated outside the horizon, 
we also take an average over the unitary operators of the
remaining black hole $U_B$. Note that $U_B$ is different from the $U$,
and does not influence the radiation outside the horizon. With this
physical picture, the fidelity relations Eq.~(\ref{APfidelity}) in the manuscript may be written

\begin{eqnarray}
	\langle F  \rangle_{U_B} \geq 1- \frac{1}{2} \sqrt{n} \int_{U_B}  \left\| \text{tr}_{\bar n}
	\Bigl(U_B \text{tr}_R(\rho_U) {U_B}^\dagger \Bigr)
	-\frac{\mathbb{1}_n}{n} \right\|_2 d U_B  . 
	\label{fidelity3} 
\end{eqnarray}
Using similar analysis to that used in Appendix C, the 2-norm in
Eq.~(\ref{fidelity3}) may be calculated as
\begin{eqnarray}
&&\int_{U_B} \left\| \text{tr}_{\bar n} \Bigl(U_B \text{tr}_R(\rho_U)
{U_B}^\dagger \Bigr)-\frac{\mathbb{1}_n}{n} \right\|^2_2 d U_B  \nonumber \\
&=& \int_{U_B} \text{tr}_{n}\left[\text{tr}_{\bar n}
\Bigl(U_B \text{tr}_R(\rho_U) {U_B}^\dagger \Bigr) \text{tr}_{\bar n}
\Bigl(U_B \text{tr}_R(\rho_U) {U_B}^\dagger \Bigr) -\frac{2}{n}
\text{tr}_{\bar n} \Bigl(U_B \text{tr}_R(\rho_U) {U_B}^\dagger \Bigr)
+\frac{\mathbb{1}_n}{n^{2}}\right] d U_B  \nonumber \\
&=& \int_{U_B} \text{tr}_{n}\left[\text{tr}_{\bar n} \Bigl(U_B
\text{tr}_R(\rho_U) {U_B}^\dagger \Bigr) \text{tr}_{\bar n}
\Bigl(U_B \text{tr}_R(\rho_U) {U_B}^\dagger \Bigr) \right] d U_B
 - \frac{2}{n} \int_{U_B} \text{tr}_{\bar R} \Bigl(U_B \text{tr}_R(\rho_U)
{U_B}^\dagger \Bigr) d U_B+\frac{1}{n}  \nonumber \\
&=& \int_{U_B} \text{tr}_{n,n}\left[\text{tr}_{\bar n , \bar n}
\Bigl(U_B \text{tr}_R(\rho_U) {U_B}^\dagger \Bigr) \otimes \Bigl(U_B
\text{tr}_R(\rho_U) {U_B}^\dagger \Bigr) \text{SWAP}_{n,n}
 \right] d U_B  - \frac{2}{n} \int_{U_B} \text{tr}_{\bar R}
( \text{tr}_R(\rho_U) ) d U_B+\frac{1}{n}    \nonumber \\ 
&=& \int_{U_B} \operatorname{tr} \left[ (U_B \otimes U_B)
\Bigl(\text{tr}_R(\rho_U) \otimes \text{tr}_R(\rho_U) \Bigr)
({U_B}^\dagger \otimes {U_B}^\dagger) \Bigl(\text{SWAP}_{n,n}
\otimes \mathbb{1}_{\bar n, \bar n} \Bigr) \right] d U_B
-\frac{1}{n} \nonumber \\
&=& \text{tr} \Biggl[ \Bigl(\text{tr}_R(\rho_U) \otimes
\text{tr}_R(\rho_U) \Bigr) \int_{U_B}  \left[({U_B}^\dagger
\otimes {U_B}^\dagger) \Bigl(\text{SWAP}_{n,n}
\otimes \mathbb{1}_{\bar n, \bar n} \Bigr)  (U_B \otimes U_B)
 \right] d U_B \Biggr]-\frac{1}{n}, 
\label{sec4result1}
\end{eqnarray}
where $\| A\|^2_2 = \text{tr}(A^\dagger A)$ is applied to the first step,
and $\text{tr}(AB) = \text{tr}( A \otimes B \; \text{SWAP}_{A,B} )$
is applied in moving from the third to the fourth line. Applying the
Schur-Weyl duality (see Appendix B) to Eq.~(\ref{sec4result1}) yields
\begin{eqnarray}
&=& \frac{1}{2} \text{tr} \Biggl[ \Bigl(\text{tr}_R(\rho_U) \otimes
\text{tr}_R(\rho_U) \Bigr)  \left(\frac{\frac{N_B}{n}+n}{N_B+1}
(\mathbb{1}_{N_B,N_B}+ \text{SWAP}_{N_B,N_B})
+  \frac{\frac{N_B}{n}-n}{N_B-1}(\mathbb{1}_{N_B,N_B}
- \text{SWAP}_{N_B,N_B}) \right) \Biggr]- \frac{1}{n} \nonumber \\
&=& \text{tr} \Biggl[ \Bigl(\text{tr}_R(\rho_U) \otimes
\text{tr}_R(\rho_U) \Bigr)   \left(\frac{\frac{(N_B)^2}{n}-n}{(N_B)^2-1}
\mathbb{1}_{N_B, N_B} + \frac{N_Bn-\frac{N_B}{n}}{(N_B)^2-1}
\text{SWAP}_{N_B, N_B} \right) \Biggr] - \frac{1}{n} \nonumber \\
&=& \frac{\frac{(N_B)^2}{n}-n}{(N_B)^2-1}
\Bigl(\text{tr}\bigl(\text{tr}_R(\rho_U)\bigr)\Bigr)^2
+ \frac{N_Bn-\frac{N_B}{n}}{(N_B)^2-1} \operatorname{tr}
\Bigr(\bigl(\text{tr}_R(\rho_U) \bigr)^2 \Bigr) - \frac{1}{n} \nonumber \\
&=& \frac{\frac{(N_B)^2}{n}-n}{(N_B)^2-1} (\text{tr}(\rho_U))^2
+ \frac{N_Bn-\frac{N_B}{n}}{(N_B)^2-1} \operatorname{tr}
\Bigr(\bigl(\text{tr}_R(\rho_U)\bigr)^2 \Bigr) - \frac{1}{n}  .
\label{sec4result2}
\end{eqnarray}
Since the interior and radiations of the black hole are both random density matrices, the term $\text{tr} \bigr((\text{tr}_R(\rho_U))^2 \bigr)$ in Eq.~(\ref{sec4result2}) may be calculated 
\begin{eqnarray}
    \left< \text{tr} \bigr((\text{tr}_R(\rho_U))^2 \bigr) \right>   
    &=& \int_U \text{tr} \bigr((\text{tr}_R(\rho_U))^2 \bigr) dU = \int_U \text{tr} \Bigl[ \text{tr}_R(\rho_U) \otimes \text{tr}_R(\rho_U) \text{SWAP}_{\bar{R},\bar{R}} \Bigr] dU  \nonumber \\
	&=&  \int_U \text{tr} \Bigl[\rho_U \otimes \rho_U \text{SWAP}_{\bar{R},\bar{R}} \Bigr] dU  =  \int_U \text{tr} \Bigl[(U \otimes U) (\rho \otimes \rho) (U^\dagger \otimes U^\dagger) \text{SWAP}_{\bar{R},\bar{R}} \Bigr]  dU  \nonumber \\
	&=& \text{tr} \biggl[ (\rho \otimes \rho)  \int_U (U^\dagger \otimes U^\dagger) \text{SWAP}_{\bar{R},\bar{R}} (U \otimes U) dU  \biggr] \nonumber \\
	&=&\frac{1}{2} \text{tr} \biggl[ (\rho \otimes \rho)  \Bigl( \frac{N/\bar{R}+\bar{R}}{N+1} (\mathbb{1}_{N,N}+ \text{SWAP}_{N,N})
+  \frac{N/\bar{R}-\bar{R}}{N-1}(\mathbb{1}_{N,N}
- \text{SWAP}_{N,N})  \Bigr) \biggr]  \nonumber \\
	&=& \text{tr} \biggl[ (\rho \otimes \rho)  \Bigl( \frac{NR-N_B}{N^2-1} \mathbb{1}_{N,N} + \frac{N N_B -R }{N^2-1} \text{SWAP}_{N,N}   \Bigr) \biggr] \nonumber \\
	&=&  \frac{NR-N_B}{N^2-1}  (\text{tr}(\rho ))^2 + \frac{N N_B -R }{N^2-1} \text{tr}(\rho^2 ) = \frac{NR-N_B}{N^2-1} + \frac{N N_B -R }{N^2-1} \nonumber \\
	&=& \frac{N_B+R}{N+1}  ,
	\label{ExtraAvera}
\end{eqnarray}
where the Schur-Weyl duality is used in moving from the third to the fourth line, $\bar{R}=N_B$ and $N_B=N/R$  are used in moving from the fourth to the fifth line, and $(\text{tr}(\rho ))^2 = \text{tr}(\rho^2 ) =1$ is used in the sixth line.

Applying Eq.~(\ref{ExtraAvera}) and $(\text{tr}(\rho_U))^2=1$ to Eq.~(\ref{sec4result2}) yields
\begin{eqnarray}
	\!\!\int_{U_B} \!\left\| \text{tr}_{\bar n} \Bigl(U_B \text{tr}_R(\rho_U)
	{U_B}^\dagger \Bigr)-\frac{\mathbb{1}_n}{n} \right\|^2_2 \!\!d U_B \! &=&
	\frac{\frac{(N_B)^2}{n}-n}{(N_B)^2-1} + \frac{N_Bn-\frac{N_B}{n}}{(N_B)^2-1} \frac{N_B+R}{N+1} - \frac{1}{n} \nonumber \\ &=&  \frac{(N_B^2-n^2)(N+1)+(N_B n^2-N_B)(N_B+R)-(N_B^2-1)(N+1)}{((N_B)^2-1)(N+1) n} \nonumber \\
	&=& \frac{N^2_B n^2 + N_B R n^2 - N n^2 -N^2_B - N_B R + N - n^2 + 1 }{((N_B)^2-1)(N+1) n}  \nonumber \\
	&=& \frac{N^2_B n^2 -N^2_B - n^2 + 1 }{((N_B)^2-1)(N+1) n} = \frac{n^2-1}{(N+1) n} ,
\label{sec4result3}
\end{eqnarray}
where $N=N_B R$ is used in moving from the second to the third line.

{  Since $N \geq n \geq 1$, Eq.~(\ref{sec4result3}) } may be further simplified as
\begin{eqnarray}
\int_{U_B} \!\left\| \text{tr}_{\bar n} \Bigl(U_B \text{tr}_R(\rho_U)
{U_B}^\dagger \Bigr)-\frac{\mathbb{1}_n}{n} \right\|^2_2 \!\!d U_B <  \frac{n}{N} .
	\label{Eresult1}
\end{eqnarray}

Inserting (\ref{Eresult1}) into Eq.~(\ref{fidelity3}) yields
\begin{eqnarray}
		\langle F  \rangle_{U_B} \geq 1- \frac{1}{2} \frac{n}{\sqrt{N}}.
\end{eqnarray}

{ Similarly to Eq.~(\ref{BHFideBound1}), we have
\begin{equation}
	\text{prob}\bigl(F \le 1 - \delta F_{\text{deviation}}\bigr)
	\le \frac{n}{\delta F_{\text{deviation}}  2\sqrt{N}} .
	\label{BHFideBound2}
\end{equation}
}

\section*{Appendix F}

Now we consider a black hole that begins
from a generic quantum state $\rho_0$, with Hilbert space dimension $N$.
The local subsystem observed by the infalling observer with dimension $n$ may be
written $\text{tr}_{\bar{n}} (\rho_{U})$. The 2-norm in the fidelity relation Eq.~(\ref{APfidelity}) in the manuscript may be calculated 
\begin{eqnarray}
\int_U \left\| \text{tr}_{\bar{n}} (\rho_U)-\frac{\mathbb{1}_n}{n}
\right\|^2_2 d U &=& \int_U \left\| \text{tr}_{\bar{n}} (U \rho_0
U^\dagger)-\frac{\mathbb{1}_n}{n} \right\|^2_2 d U \nonumber \\
&=& \int_U \operatorname{tr}_{n}\left[\operatorname{tr}_{\bar n}
(U \rho_0 U^\dagger) \operatorname{tr}_{\bar n}
(U \rho_0 U^\dagger) -\frac{2}{n}
\operatorname{tr}_{\bar n}(U \rho_0 U^\dagger)
+\frac{\mathbb{1}_n}{n^{2}}\right] d U \nonumber \\
 &=& \int_U \operatorname{tr}_{n}\left[\operatorname{tr}_{\bar n}
(U \rho_0 U^\dagger) \operatorname{tr}_{\bar n}
(U \rho_0 U^\dagger) \right] d U -\frac{1}{n} \nonumber \\
 &=& \int_U \operatorname{tr}_{n,n}\left[\operatorname{tr}_{\bar n , \bar n}
(U \rho_0 U^\dagger) \otimes (U \rho_0
U^\dagger) \, \text{SWAP}_{n,n} \right] d U -\frac{1}{n} \nonumber \\
 &=& \int_U \operatorname{tr} \left[ (U \otimes U) \Bigl(
\rho_0 \otimes \rho_0 \Bigr)
( U^\dagger \otimes U^\dagger ) \Bigl(\text{SWAP}_{n,n}
\otimes \mathbb{1}_{\bar n, \bar n} \Bigr) \right] d U -\frac{1}{n} , \nonumber \\ 
 &=& \operatorname{tr} \Biggl[ (\rho_0 \otimes \rho_0)  \int_U  \left( (U^\dagger \otimes U^\dagger) \Bigl(\text{SWAP}_{n,n}
\otimes \mathbb{1}_{\bar n, \bar n} \Bigr) (U \otimes U)
 \right) d U \Biggr] -\frac{1}{n}  
\label{R1}
\end{eqnarray}
{where $\operatorname{tr}_{n} (\operatorname{tr}_{\bar n}(U \rho_0 U^\dagger) ) = \operatorname{tr}(U \rho_0 U^\dagger)=1$ is applied to the second line.}
Applying the Schur-Weyl duality to Eq.~(\ref{R1}) yields
\begin{eqnarray}
\int_U \left\| \text{tr}_{\bar{n}} (\rho_U)-\frac{\mathbb{1}_n}{n}
\right\|^2_2 d U  &=& \frac{1}{2} \operatorname{tr} \Biggl[ (\rho_0 \otimes \rho_0) 
\left(\frac{\frac{N}{n}+n}{N+1}(\mathbb{1}_{N,N}+ \text{SWAP}_{N,N})
+  \frac{\frac{N}{n}-n}{N-1}(\mathbb{1}_{N,N}
- \text{SWAP}_{N,N}) \right) \Biggr]-
\frac{1}{n} \nonumber \\
 &=& \operatorname{tr} \Biggl[ (\rho_0 \otimes \rho_0) 
\left(\frac{\frac{N^2}{n}-n}{N^2-1} \mathbb{1}_{N,N}
+ \frac{Nn-\frac{N}{n}}{N^2-1} \text{SWAP}_{N,N} \right) \Biggr]
- \frac{1}{n} \nonumber \\
 &=& \frac{\frac{N^2}{n}-n}{N^2-1} (\operatorname{tr} \rho_0 )^2
+ \frac{Nn-\frac{N}{n}}{N^2-1} \operatorname{tr} (\rho_0^2 )
- \frac{1}{n} = \frac{\frac{N^2}{n}-n}{N^2-1}
+ \frac{Nn-\frac{N}{n}}{N^2-1} \operatorname{tr} (\rho_0^2 ) - \frac{1}{n}  \nonumber \\
 &=& \frac{Nn^2-N}{(N^2-1)n} \operatorname{tr} (\rho_0^2 ) - \frac{n^2-1}{(N^2-1)n} = \frac{(N \operatorname{tr} (\rho_0^2 ) -1)(n^2-1)}{(N^2-1)n}  ,
 \label{R2}
\end{eqnarray}
where $(\operatorname{tr} \rho_0 )=1$ is
used in the third line because the entire black hole is a pure quantum state. 

{ Note that as $n\le N$, then $ N^2 n^2 -N^2 \le N^2 n^2 -n^2$ and hence
$N^2(n^2-1) \le n^2(N^2-1)$.
Further taking  $1< N$, and using Eq.~(\ref{R2}) and Jensen's inequality
we find, }
\begin{eqnarray}
	\int_U \left\| \text{tr}_{\bar{n}} (\rho_U)
	-\frac{\mathrm{I}_n}{n} \right\|_2 d U
	\leq \frac{\sqrt{n(N \operatorname{tr} (\rho_0^2 ) -1)}}{N} .
	\label{resultR1}
\end{eqnarray}

Inserting Eqs.~(\ref{resultR1}) into the fidelity relations Eq.~(\ref{APfidelity}) yields
\begin{eqnarray}
		\langle F  \rangle_U \geq 1 - \frac{n}{2N} \sqrt{N \operatorname{tr}
		(\rho_0^2 ) -1}  .
\end{eqnarray}

{  
Similarly to the pure state scenario, the average fidelity suggests that the probability, $\text{prob}$, of the fidelity deviating by $\delta F_{\text{deviation}}$ from 1 is extremely small, and $\text{prob}$ should obey
\begin{equation}
	\text{prob}\bigl(F \le 1 - \delta F_{\text{deviation}}\bigr)
	\le \frac{n \sqrt{N \operatorname{tr}
	(\rho_0^2 ) -1}}{\delta F_{\text{deviation}} 2N} .
	\label{BHFideBound3}
\end{equation}
}

\section*{Appendix G}

Similarly to Appendix E, we can also consider black holes originating from {  a generic black hole that has } non-negligible radiation. If we still assume the initial dimension of the entire black hole state
to be $N$ and the dimension of the radiation is $R$, then the dimension
of the remaining black hole $N_B$ will equal $N/R$. The quantum state of
the remaining black hole may be represented by $\text{tr}_R (\rho_U)$.
The key step in our calculation of the 2-norm is to take an average over
all possible unitary operators, since black holes are fast scramblers.
Now, with part of the initial state radiated outside the horizon, 
we first take an average over the unitary operators of the
remaining black hole $U_B$. Here $U_B$ is different from the $U$,
and does not influence the radiation outside the horizon. With this
physical picture, the fidelity will have a range of
\begin{eqnarray}
	\langle F  \rangle_{U_B} \geq 1- \frac{1}{2} \sqrt{n} \int_{U_B} \left\| \text{tr}_{\bar n}
\Bigl(U_B \text{tr}_R(\rho_U) {U_B}^\dagger \Bigr)
-\frac{\mathbb{1}_n}{n} \right\|_2 d U_B .
	\label{Genefidelity1}
\end{eqnarray}

The 2-norm in Eq.~(\ref{Genefidelity1}) may be calculated
\begin{eqnarray}
&&\int_{U_B} \left\| \text{tr}_{\bar n} \Bigl(U_B \text{tr}_R(\rho_U)
{U_B}^\dagger \Bigr)-\frac{\mathbb{1}_n}{n} \right\|^2_2 d U_B  \nonumber \\
&=& \int_{U_B} \text{tr}_{n}\left[\text{tr}_{\bar n}
\Bigl(U_B \text{tr}_R(\rho_U) {U_B}^\dagger \Bigr) \text{tr}_{\bar n}
\Bigl(U_B \text{tr}_R(\rho_U) {U_B}^\dagger \Bigr) -\frac{2}{n}
\text{tr}_{\bar n} \Bigl(U_B \text{tr}_R(\rho_U) {U_B}^\dagger \Bigr)
+\frac{\mathbb{1}_n}{n^{2}}\right] d U_B  \nonumber \\
&=& \int_{U_B} \text{tr}_{n}\left[\text{tr}_{\bar n} \Bigl(U_B
\text{tr}_R(\rho_U) {U_B}^\dagger \Bigr) \text{tr}_{\bar n}
\Bigl(U_B \text{tr}_R(\rho_U) {U_B}^\dagger \Bigr) \right] d U_B
 - \frac{2}{n} \int_{U_B} \text{tr}_{\bar R} \Bigl(U_B \text{tr}_R(\rho_U)
{U_B}^\dagger \Bigr) d U_B+\frac{1}{n}  \nonumber \\
&=& \int_{U_B} \text{tr}_{n,n}\left[\text{tr}_{\bar n , \bar n}
\Bigl(U_B \text{tr}_R(\rho_U) {U_B}^\dagger \Bigr) \otimes \Bigl(U_B
\text{tr}_R(\rho_U) {U_B}^\dagger \Bigr) \text{SWAP}_{n,n}
 \right] d U_B  - \frac{2}{n} \int_{U_B} \text{tr}_{\bar R}
( \text{tr}_R(\rho_U) ) d U_B+\frac{1}{n}    \nonumber \\ 
&=& \int_{U_B} \operatorname{tr} \left[ (U_B \otimes U_B)
\Bigl(\text{tr}_R(\rho_U) \otimes \text{tr}_R(\rho_U) \Bigr)
({U_B}^\dagger \otimes {U_B}^\dagger) \Bigl(\text{SWAP}_{n,n}
\otimes \mathbb{1}_{\bar n, \bar n} \Bigr) \right] d U_B
-\frac{1}{n} \nonumber \\
&=& \text{tr} \Biggl[ \Bigl(\text{tr}_R(\rho_U) \otimes
\text{tr}_R(\rho_U) \Bigr) \int_{U_B}  \left[({U_B}^\dagger
\otimes {U_B}^\dagger) \Bigl(\text{SWAP}_{n,n}
\otimes \mathbb{1}_{\bar n, \bar n} \Bigr)  (U_B \otimes U_B)
 \right] d U_B \Biggr]-\frac{1}{n}, 
\label{Generesult1}
\end{eqnarray}
where $\| A\|_2 = \text{tr}(A^\dagger A)$ is applied to the first step,
and $\text{tr}(AB) = \text{tr}( A \otimes B \; \text{SWAP}_{A,B} )$
is applied in moving from the third to the fourth line. Applying the
Schur-Weyl duality in Appendix B to Eq.~(\ref{Generesult1}) yields
\begin{eqnarray}
&&\int_{U_B} \left\| \text{tr}_{\bar n} \Bigl(U_B \text{tr}_R(\rho_U)
{U_B}^\dagger \Bigr)-\frac{\mathbb{1}_n}{n} \right\|^2_2 d U_B \nonumber \\
&=& \frac{1}{2} \text{tr} \Biggl[ \Bigl(\text{tr}_R(\rho_U) \otimes
\text{tr}_R(\rho_U) \Bigr)  \left(\frac{\frac{N_B}{n}+n}{N_B+1}
(\mathbb{1}_{N_B,N_B}+ \text{SWAP}_{N_B,N_B})
+  \frac{\frac{N_B}{n}-n}{N_B-1}(\mathbb{1}_{N_B,N_B}
- \text{SWAP}_{N_B,N_B}) \right) \Biggr]- \frac{1}{n} \nonumber \\
&=& \text{tr} \Biggl[ \Bigl(\text{tr}_R(\rho_U) \otimes
\text{tr}_R(\rho_U) \Bigr)   \left(\frac{\frac{(N_B)^2}{n}-n}{(N_B)^2-1}
\mathbb{1}_{N_B, N_B} + \frac{N_Bn-\frac{N_B}{n}}{(N_B)^2-1}
\text{SWAP}_{N_B, N_B} \right) \Biggr] - \frac{1}{n} \nonumber \\
&=& \frac{\frac{(N_B)^2}{n}-n}{(N_B)^2-1}
\Bigl(\text{tr}\bigl(\text{tr}_R(\rho_U)\bigr)\Bigr)^2
+ \frac{N_Bn-\frac{N_B}{n}}{(N_B)^2-1} \operatorname{tr}
\Bigr(\bigl(\text{tr}_R(\rho_U) \bigr)^2 \Bigr) - \frac{1}{n} \nonumber \\
&=& \frac{\frac{(N_B)^2}{n}-n}{(N_B)^2-1} (\text{tr}(\rho_U))^2
+ \frac{N_Bn-\frac{N_B}{n}}{(N_B)^2-1} \operatorname{tr}
\Bigr(\bigl(\text{tr}_R(\rho_U)\bigr)^2 \Bigr) - \frac{1}{n} \nonumber \\
&=& \frac{\frac{(N_B)^2}{n}-n}{(N_B)^2-1} + \frac{N_Bn-\frac{N_B}{n}}{(N_B)^2-1} \operatorname{tr}
\Bigr(\bigl(\text{tr}_R(\rho_U)\bigr)^2 \Bigr) - \frac{1}{n}  ,
\label{Generesult2}
\end{eqnarray}
where we have used $(\text{tr}(\rho_U))^2=1$ in the last step.

Since black holes are believed to be fast quantum scrambler, a black hole should already finish its quantum {  scrambling before it radiates } too much state away. Thus. the term $\text{tr} \bigr((\text{tr}_R(\rho_U))^2 \bigr)$ in Eq.~(\ref{Generesult2}) may be calculated 
\begin{eqnarray}
    \left< \text{tr} \bigr((\text{tr}_R(\rho_U))^2 \bigr) \right>   
    &=& \int_U \text{tr} \bigr((\text{tr}_R(\rho_U))^2 \bigr) dU = \int_U \text{tr} \Bigl[ \text{tr}_R(\rho_U) \otimes \text{tr}_R(\rho_U) \text{SWAP}_{\bar{R},\bar{R}} \Bigr] dU  \nonumber \\
	&=&  \int_U \text{tr} \Bigl[\rho_U \otimes \rho_U \text{SWAP}_{\bar{R},\bar{R}} \Bigr] dU  =  \int_U \text{tr} \Bigl[(U \otimes U) (\rho \otimes \rho) (U^\dagger \otimes U^\dagger) \text{SWAP}_{\bar{R},\bar{R}} \Bigr]  dU  \nonumber \\
	&=& \text{tr} \biggl[ (\rho \otimes \rho)  \int_U (U^\dagger \otimes U^\dagger) \text{SWAP}_{\bar{R},\bar{R}} (U \otimes U) dU  \biggr] \nonumber \\
	&=&\frac{1}{2} \text{tr} \biggl[ (\rho \otimes \rho)  \Bigl( \frac{N/\bar{R}+\bar{R}}{N+1} (\mathbb{1}_{N,N}+ \text{SWAP}_{N,N})
+  \frac{N/\bar{R}-\bar{R}}{N-1}(\mathbb{1}_{N,N}
- \text{SWAP}_{N,N})  \Bigr) \biggr]  \nonumber \\
	&=& \text{tr} \biggl[ (\rho \otimes \rho)  \Bigl( \frac{NR-N_B}{N^2-1} \mathbb{1}_{N,N} + \frac{N N_B -R }{N^2-1} \text{SWAP}_{N,N}   \Bigr) \biggr] \nonumber \\
	&=&  \frac{NR-N_B}{N^2-1}  (\text{tr}(\rho ))^2 + \frac{N N_B -R }{N^2-1} \text{tr}(\rho^2 ) \nonumber \\
	&=& \frac{NR-N_B+ (N N_B -R)\text{tr}(\rho^2 )}{N^2-1},
	\label{GeneExtraAvera}
\end{eqnarray}
where the Schur-Weyl duality is used in moving from the third to the fourth line, $\bar{R}=N_B$ and $N_B=N/R$  are used in moving from the fourth to the fifth line, and $(\text{tr}(\rho ))^2 = \text{tr}(\rho^2 ) =1$ is applied to the sixth line.

Applying Eq.~(\ref{GeneExtraAvera}) to Eq.~(\ref{Generesult2}) yields
\begin{eqnarray}
	&&\int_{U_B} \left\| \text{tr}_{\bar n} \Bigl(U_B \text{tr}_R(\rho_U)
{U_B}^\dagger \Bigr)-\frac{\mathbb{1}_n}{n} \right\|^2_2 d U_B \nonumber \\
&=&\frac{\frac{(N_B)^2}{n}-n}{(N_B)^2-1} + \frac{N_Bn-\frac{N_B}{n}}{(N_B)^2-1} \frac{NR-N_B+ (N N_B -R)\text{tr}(\rho^2 )}{N^2-1} - \frac{1}{n} \nonumber \\
	&=&  \frac{1-n^2}{((N_B)^2-1)n} + \frac{N_Bn^2 - N_B }{((N_B)^2-1)n} \frac{NR-N_B+ (N N_B -R)\text{tr}(\rho^2 )}{N^2-1} \nonumber \\
	&=& \frac{(1-n^2)(N^2-1)}{((N_B)^2-1)(N^2-1)n} + \frac{N^2(n^2 - 1)-N_B^2(n^2-1) + (N_Bn^2 - N_B )(N N_B -R)\text{tr}(\rho^2 ) }{((N_B)^2-1)(N^2-1)n} \nonumber \\
	&=& \frac{(n^2 - 1)-(N_B)^2(n^2-1) + N((N_B)^2 -1)(n^2 - 1)\text{tr}(\rho^2 ) }{((N_B)^2-1)(N^2-1)n}  \nonumber \\
	&=& \frac{-(n^2 - 1)+ N(n^2 - 1)\text{tr}(\rho^2 ) }{(N^2-1)n}  \nonumber \\
	&=& \frac{(N\text{tr}(\rho^2 ) - 1)(n^2 - 1) }{(N^2-1)n}  .
	\label{N2norm1}
\end{eqnarray}
where $N=N_B R$ is used in moving from the third to the four line line.
{As $n\le N$, then
$N^2(n^2-1) \le n^2(N^2-1)$.
Further taking  $1<N$, and using Eq.~(\ref{N2norm1}) and Jensen's inequality
we find }
\begin{eqnarray}
	\int_{U_B} \left\| \text{tr}_{\bar n} \Bigl(U_B \text{tr}_R(\rho_U)
{U_B}^\dagger \Bigr)-\frac{\mathbb{1}_n}{n} \right\|^2_2 d U_B = \frac{(N\text{tr}(\rho^2 ) - 1)(n^2 - 1) }{(N^2-1)n} \leq
\frac{(N\text{tr}(\rho^2 ) - 1)n}{N^2}  .
	\label{N2norm2}
\end{eqnarray}
Applying Eq.~(\ref{N2norm2}) to Eq.~(\ref{Genefidelity1}) yields
\begin{eqnarray}
	\langle F  \rangle_{U_B} \geq 1- \frac{n}{2N} \sqrt{N\text{tr}(\rho^2 ) - 1} .
	\label{Genefidelity2}
\end{eqnarray}
{   Similarly to Eq.~(\ref{BHFideBound3}), we have
	\begin{equation}
		\text{prob}\bigl(F \le 1 - \delta F_{\text{deviation}}\bigr)
		\le \frac{n \sqrt{N \operatorname{tr}
				(\rho_0^2 ) -1}}{\delta F_{\text{deviation}} 2N} .
		\label{BHFideBound4}
	\end{equation}
}

\end{widetext}

\end{document}